\documentclass[%
twocolumn,
superscriptaddress,
 amsmath,amssymb, aps, prc, floatfix,
]{revtex4-1}

\usepackage{amsmath}
\usepackage{amssymb}
\usepackage{amsfonts}
\usepackage{comment}

\usepackage{xcolor}

\usepackage{dsfont}

\usepackage{braket}
\usepackage{graphicx}% Include figure files
\usepackage{dcolumn}% Align table columns on decimal point
\usepackage{bm}% bold math

\newcommand{\half}[0]
	{
	  \frac{1}{2} 
	}

\newcommand{\ocal}[0]
	{
	  { \mathcal{O} }_{ \lambda } 
	}

\newcommand{\an}[0]
	{
	  { a }_{ \nu } 
	}

\newcommand{\bn}[0]
	{
	  { b }_{ \nu } 
	}

\begin{document}

\preprint{}

\title{Magnetic dipole ${\gamma}$-ray strength functions in the crossover from spherical to deformed neodymium isotopes} 

\author{A. Mercenne}
\affiliation{Center for Theoretical Physics, Sloane Physics Laboratory, Yale University, New Haven, Connecticut 06520, USA}
\affiliation{ Department of Physics and Astronomy, Louisiana State University, Baton Rouge, Louisiana 70803, USA }
\author{P. Fanto}
\affiliation{Center for Theoretical Physics, Sloane Physics Laboratory, Yale University, New Haven, Connecticut 06520, USA}
\author{W. Ryssens}
\affiliation{Institut d'Astronomie et d'Astrophysique, Universit{\'e} Libre de Bruxelles,
Campus de la Plaine CP 226, 1050 Brussels, Belgium}
\author{Y. Alhassid}
\affiliation{Center for Theoretical Physics, Sloane Physics Laboratory, Yale University, New Haven, Connecticut 06520, USA}

%\date{\today}% 
\begin{abstract}
We calculate the magnetic dipole $\gamma$-ray strength functions in a chain of even-mass neodymium isotopes $^{144-152}$Nd in the framework of the configuration-interaction (CI) shell model. We infer the strength function by applying the maximum entropy method (MEM)  to the exact imaginary-time response function calculated with the shell-model Monte Carlo (SMMC) method. The success of the MEM depends on the choice of a good strength function as a prior distribution. We investigate two choices for the prior strength function: the static path approximation (SPA) and the quasiparticle random-phase approximation (QRPA). We find that the QRPA is a better approximation at low temperatures (i.e., near the ground state), while the SPA is a better choice at finite temperatures.
We identify a low-energy enhancement (LEE) in the MEM deexcitation $M1$ strength functions of the even-mass neodymium isotopes and compare with recent experimental results for the total deexcitation $\gamma$-ray strength functions.  The LEE is already seen in the SPA strength function but not in the QRPA strength function, indicating the importance of large-amplitude static fluctuations around the mean field in reproducing the LEE.   Our method is currently the only one which can reproduce LEE in heavy open-shell nuclei where conventional CI shell model calculations are prohibited.
With the onset of deformation as number of neutrons increases along the chain of neodymium isotopes, we observe that some of the LEE strength transfers to a low-energy excitation, which we interpret as a finite-temperature ``scissors'' mode.  We also observe a finite-temperature spin-flip mode.

\end{abstract}

\pacs{Valid PACS appear here}% PACS, the Physics and Astronomy
\maketitle

\section{Introduction}

 The process of creating new elements through the rapid capture of neutrons, known as r-process nucleosynthesis, can be studied by modeling neutron capture rates in compound nucleus reactions.  These neutron capture rates are calculated within the Hauser-Feshbach theory~\cite{hauser_1952,koonin_2012}, which requires as input the  $\gamma$-ray strength function (${\gamma}$SF)~\cite{bartholomev_1973}. By accurately determining the ${\gamma}$SF, we can improve our ability to model r-process nucleosynthesis~\cite{mumpower_2016} and gain a better understanding of the origins and abundances of the elements in the universe.
 
 In heavy nuclei, the magnetic dipole ($M1$) ${\gamma}$SF exhibits several excitation modes that are typically observed in the low-energy tail of the giant dipole resonance. These modes include a ``scissors'' mode and a spin-flip resonance~\cite{bohle_1984,heyde_2010}.  In strongly deformed nuclei the orbital scissors mode at lower energies is clearly separated from a resonance-like structure composed of proton and neutron spin-flip excitations at higher energies~\cite{heyde_2010}.   The inclusion of the scissors mode in the $M1$ $\gamma$SF improves the accuracy of neutron radiative capture rates in Hauser-Feshbach calculations~\cite{mumpower_2017}.  
 
A low-energy enhancement (LEE), characterized by an increase in strength  at low ${\gamma}$-ray energies, has been observed in the ${\gamma}$SFs of several mid-mass nuclei~\cite{voinov_2004,guttormsen_2005,wiedeking_2012,larsen_2013,larsen_2018} and several heavier lanthanide  nuclei~\cite{kheswa_2015,navqi_2019,simon_2016,guttormsen_2022}. The LEE was shown to be dipole in nature~\cite{jones_2018}.
If the LEE persists in heavy, neutron-rich nuclei, it would significantly increase the radiative neutron capture cross sections of nuclei near the neutron drip line~\cite{larsen_2010}. 

Configuration-interaction (CI) shell-model calculations~\cite{loens_2012,schwengner_2013,brown_2014,schwengner_2017,sieja_2017,karampagia_2017,sieja_2018} suggested that the LEE originates in the $M1$  $\gamma$SF, although relativistic QRPA calculations~\cite{litvinova_2013} proposed the LEE is of $E1$ origin.  CI shell model calculations were carried out  for nuclei in different major shells and concluded that the LEE is more pronounced near shell closure~\cite{mitdbo_2018}.  

Conventional CI shell-model methods in heavy open-shell nuclei are  prohibited  because of the combinatorial increase of the dimensionality of the many-particle model space with number of valence nucleons and/or number of valence orbitals.   The shell-model Monte Carlo (SMMC) method~\cite{johnson_1992,alhassid_1994,nakada_1997,nakada_1998,ormand_1997,langanke_1998,alhassid_1999,alhassid_2007,alhassid_2008,alhassid_2017_book} uses auxiliary-field Monte Carlo techniques within the CI shell-model framework to enable exact (within statistical errors) calculations of thermal observables in heavy nuclei.   However, in SMMC we cannot calculate directly the finite-temperature $\gamma$SF, but only its Laplace transform, the imaginary-time response function. The calculation of the $\gamma$SF requires an analytic continuation which is carried out in the maximum entropy method (MEM)~\cite{gubernatis_1991, jarrell_1996,gubernatis_book}.
 
 The success of the MEM depends on a good choice for a prior strength function at finite temperature. In Ref.~\cite{fanto_2024}, the static path approximation plus random-phase approximation (SPA+RPA) was used to calculate a prior for the $M1$ $\gamma$SF for a pairing plus quadrupole interaction, and a LEE was identified in a chain of samarium isotopes.  However, for more general interactions such as the interactions used in SMMC applications to lanthanides~\cite{ozen_2013}, the SPA+RPA becomes time consuming.   Here we investigate two other choices for the prior: the static path approximation (SPA)~\cite{muhlschlegel_1972, zingman_1984, lauritzen_1988, arve_1988, attias_1997} and the finite-temperature quasiparticle random phase approximation (QRPA). In these approximations we can calculate a strength function directly without the need for analytic continuation.  
 
We apply our methods to calculate the $M1$ $\gamma $SFs in a chain of even-mass neodymium isotopes $^{144-152}$Nd.  We convert these strength functions to deexcitation $M1$ strength functions and compare our results with recent experimental results for the deexcitation $\gamma$SFs in $^{144-150}$Nd~\cite{guttormsen_2022}.  We also provide a prediction of the deexcitation $M1$ $\gamma$SF in $^{152}$Nd, for which no measurements were made. 

Comparing the imaginary-time response functions of the SPA and QRPA to exact SMMC response functions, we find that the QRPA is a better approximation at low temperatures close to the ground state, but that the SPA is a significantly better approximation at finite temperature. Consequently, at finite temperature we use the SPA strength as a prior for the MEM and identify a LEE in all of the even-mass neodymium isotopes in the chain under study.  With the onset of deformation with increasing number of neutrons along the chain of neodymium isotopes, we observe that some of the LEE strength transfers to a low-energy excitation, which we interpret as a finite-temperature ``scissors'' mode.   

We also find that at finite temperature the LEE is already observed in the SPA strength function but not in the QRPA strength function.  Thus the QRPA, which is the standard method for calculating finite-temperature strength functions in nuclear astrophysics, cannot reproduce the LEE~\cite{fanto_2024}.  A similar conclusion was reached in Ref.~\cite{ frosini_2023} by comparing the finite-temperature QRPA $M1$ deexcitation strength function with exact diagonalization results in the CI shell model.  We conclude that large-amplitude static fluctuations around the mean-field state, which are included in the SPA but not in the QRPA, play an important role in reproducing the LEE.
 
\section{Theoretical framework}
\label{Formalism}

\subsection{SMMC and imaginary-time response functions}

  The SMMC method is based on the representation of the Gibbs operator ${ { e }^{ -\beta \hat{ H } } }$ (where ${ \hat{ H } }$ is the CI shell model Hamiltonian and ${ \beta=1/T }$ is the inverse temperature) as a functional integral of one-body propagators $\hat U_\sigma \equiv \hat U_\sigma(\beta,0)$. These propagators describe  non-interacting nucleons moving in external auxiliary fields $\sigma = \sigma(\tau)$, which depend on imaginary time $\tau$.
  This is formally expressed in terms of the Hubbard-Stratonovich (HS) transformation~\cite{hubbard,*stratonovich_1957,*stratonovich_1958}
  \begin{equation}
    e^{ -\beta \hat{ H } } = \int_{} D[\sigma] { G }_{ \sigma } { \hat{ U } }_{ \sigma }\;,
    \label{}
  \end{equation}
  where ${ G }_{ \sigma }$ is a Gaussian factor, and $D[\sigma]$ is a measure.
  The thermal expectation value of an observable ${ \langle \hat{ O } \rangle}=  \text{Tr}(\hat{ O } e^{-\beta \hat H})/ \text{Tr}  e^{-\beta \hat H} $ can then be written as
  \begin{equation}
    \langle \hat{ O } \rangle = \frac{ \int_{} D[\sigma] { W }_{ \sigma } { \Phi }_{ \sigma } { \langle O \rangle }_{ \sigma } }{ \int_{} D[\sigma] { W }_{ \sigma } { \Phi }_{ \sigma } } \;,
    \label{thermal_obs}
  \end{equation}
  where ${ { W }_{ \sigma } = { G }_{ \sigma } |\text{Tr} { \hat{ U } }_{ \sigma }| }$ is a positive-definite weight, ${ { \Phi }_{ \sigma } = \text{Tr\,} { \hat{ U } }_{ \sigma } / |\text{Tr\,} { \hat{ U } }_{ \sigma }| }$ is the Monte Carlo sign function, and ${ { \langle O \rangle }_{ \sigma } = \text{Tr}(\hat{ O } { \hat{ U } }_{ \sigma })/ \text{Tr} { \hat{ U } }_{ \sigma } }$.
  
  In SMMC, the right-hand side of Eq.~(\ref{thermal_obs}) is calculated using stochastic sampling of uncorrelated auxiliary-field configurations ${ \sigma_k }$ according to the positive-definite distribution ${ { W }_{ \sigma } }$.   The thermal expectation of the observable $\hat O$ is then estimated by
  \begin{equation}
   \langle \hat{ O } \rangle \approx { \sum_k \langle O \rangle_{\sigma_k} \Phi_{\sigma_k} \over \sum_k \Phi_{\sigma_k}} \;.
  \end{equation}
 Thermal averages at fixed number of protons and neutrons are calculated by particle-number projection~\cite{alhassid_2017_book}. 

The imaginary-time response function of a spherical tensor $\ocal$ of rank $\lambda$ at temperature $T$ is defined by
\begin{eqnarray}\label{response-function}
R_{\mathcal{O}_\lambda}(T; \tau) & =  \langle \ocal(\tau) \cdot \ocal(0) \rangle =\sum_{\substack{\alpha_i J_i \\ \alpha_f J_f}}  \frac{e^{-\beta E_{\alpha_i J_i}}}{Z}  \nonumber \\  \times & |(\alpha_f J_f || \hat{\mathcal{O}}_\lambda|| \alpha_i J_i)|^2 e^{-\tau (E_{\alpha_f J_f} - E_{\alpha_i J_i})} \;,
\end{eqnarray}
where $\langle \ldots \rangle$ denotes a thermal average,  $ \ocal(\tau) \cdot \ocal(0)=\sum_\mu (-)^\mu \mathcal{O}_{\lambda \mu}(\tau) \mathcal{O}_{\lambda -\mu}$, and $ \mathcal{O}_{\lambda \mu}(\tau) = e^{\tau\hat H} \mathcal{O}_{\lambda \mu} e^{-\tau \hat H}$. Here  $\alpha J$ are eigenstates of the Hamiltonian $\hat H$ with spin $J$ and energy $E_{\alpha J}$. 

 Using the HS transformation, we can express this response function in the form~\cite{koonin_1997,alhassid_2017_book}
\begin{equation}\label{response-SMMC}
  { R }_{ \ocal } (T;\tau) = \frac{ \int_{} D[\sigma] W_\sigma \Phi_\sigma  { \langle \ocal (\tau) . \ocal \rangle }_{ \sigma } }{ \int_{} D[\sigma] W_\sigma\Phi_\sigma}
\end{equation}
  where ${ \ocal (\tau) = { \hat{ U } }_{ \sigma }^{ -1 } (\tau,0) \ocal { \hat{ U } }_{ \sigma } (\tau,0) }$, with $\hat U_\sigma(\tau,0)$ is the propagator between times $0$ and $\tau$ for a given configuration of the auxiliary fields $\sigma$, and the expectation value ${ { \langle \dots \rangle }_{ \sigma } }$ is taken with respect to the propagator ${ { \hat{ U } }_{ \sigma } \equiv { \hat{ U } }_{ \sigma } (\beta,0) }$. 
  As mentioned above, in SMMC we sample auxiliary-field configurations according to the weight function ${ {W}_{ \sigma } }$ and average over these samples to determine the imaginary-time response function.
  
  \subsection{Strength function and maximum entropy method}
  
  The finite-temperature strength function ${S}_{\ocal}(T;\omega)$ at temperature $T$ and transition energy $\omega$ for a tensor operator of rank $\lambda$ is defined by 
 \begin{eqnarray}\label{strength}
S_{\mathcal{O}_\lambda}(T; \omega) = \sum_{\substack{\alpha_i J_i \\ \alpha_f J_f}} &\frac{e^{-\beta E_{\alpha_i J_i}}}{Z}   |(\alpha_f J_f || \hat{\mathcal{O}}_\lambda || \alpha_i J_i )|^2 \nonumber \\ & \times\delta(\omega - E_f + E_i) \;.
\end{eqnarray} 

We note that the strength function relevant to experiments is described by the microcanonical ensemble $\delta(E_i-\hat H)$, where $E_i$ is the initial energy of the nucleus~\cite{alhassid_1990}. In the finite-temperature approach we approximate the microcanonical average by a canonical average as in Eq.~(\ref{strength}). The microcanonical ensemble is related to the canonical ensemble by an inverse Laplace transform.  In the saddle-point approximation, the temperature $T$ of the canonical ensemble is determined by the condition $E(T)=E_i$, where $E(T)=\langle \hat H\rangle$ is the average thermal energy and $E_i$ is the energy of the microcanonical ensemble. In the following we measure $E_i$ with respect to the ground-state energy, i.e., $E_i$ will denote the initial excitation energy of the nucleus.

 The response function (\ref{response-function})  is a Laplace transform of the strength function (\ref{strength})
 \begin{equation}\label{Laplace}
 R_{\mathcal{O}_\lambda}(T; \tau) = \int_{-\infty}^\infty d\omega\, e^{-\tau \omega} S_{\mathcal{O}_\lambda}(T; \omega) \;.
 \end{equation}
  Using the symmetry property of the strength function
\begin{equation}
S_{\mathcal{O}_\lambda}(T; -\omega) = e^{-\beta \omega} S_{\mathcal{O}_\lambda}(T; \omega) \;,
\end{equation}
we can rewrite (\ref{Laplace}) as an integral over non-negative values of $\omega$
  \begin{equation}\label{symmetrized_kernel}
  { R }_{ \ocal } (T;\tau) = \int_{0}^{\infty} d\omega K(\tau,\omega) { S }_{ \ocal } (T;\omega) \;,
  \end{equation}
  where  the kernel is ${ K(\tau;\omega) = { e }^{ -\tau \omega } + { e }^{ - (\beta - \tau) \omega } }$.

To determine the strength function from the response function, we have to invert ({\ref{symmetrized_kernel}) numerically which is an ill-defined problem with no unique solution.  In particular, the response function is not sensitive to the strength function at large values of $\omega$ due to the presence of an exponentially small tail in the kernel for large $\omega$.  To select an optimal strength function, we implement the maximum-entropy method (MEM)~\cite{jarrell_1996}, in which the objective function 
\begin{equation}
 \mathcal{Q} ({ S }_{ \ocal } ; \alpha) = \alpha \mathcal{S} - \frac{ 1 }{ 2 } { \chi }^{ 2 } \;
 \label{maxent_Q}
 \end{equation}
 is maximized.   In Eq.~(\ref{maxent_Q})
 \begin{equation}
    { \chi }^{ 2 } = { \left( { \bar{R} }_{ \ocal } - { R }_{ \ocal } \right) }^{ T } { \mathcal{C} }^{ -1 } \left( { \bar{R} }_{ \ocal } - { R }_{ \ocal } \right)\;,
  \end{equation}
  where ${ \bar{R} }$ and ${ \mathcal{C} }$ are the SMMC response function and its covariance matrix, respectively. 
   ${\mathcal{S} }$ is the entropy function 
 \begin{align}
    \mathcal{S} & =  - \int_{} d\omega \left( { S }_{ \ocal }(T; \omega) - { S }_{ \ocal }^{ \text{prior} } (T;\omega) \right. \nonumber \\
    & \left. - { S }_{ \ocal } (T; \omega) \ln{ \left[ { S }_{ \ocal } (T;\omega) / { S }_{ \ocal }^{ \text{prior} } (T;\omega) \right] } \right) \;,
    \label{eq_entropy}
  \end{align}
  where ${ { S }_{ \ocal }^{ \text{prior} } }$ is a suitably chosen prior for the strength function.
The MEM provides a best fit to the known SMMC response function (minimizing $\chi^2$) while at the same time keeping as close as possible to the prior strength function (maximizing $\mathcal{S}$). The coefficient $\alpha$ in the objective function (\ref{maxent_Q}) controls the relative importance between the entropy and the $\chi^2$ terms.
  
  Here we use Bryan's method~\cite{bryan_1990}, in which the final expression for the  MEM strength function is given by
\begin{equation}
    { S }_{ \ocal }^{ \text{MEM} } = \int_{} d \alpha { S }_{ \ocal }^{ \alpha } \mathcal{P} \left( \alpha | { \bar{G} }_{ \ocal } , \mathcal{C}, { S }_{ \ocal }^{ \text{prior} } \right),
    \label{}
  \end{equation}
  where ${ { S }_{ \ocal }^{ \alpha } }$ maximizes the objective function (\ref{maxent_Q}) for a given ${ \alpha }$, and  ${ \mathcal{P} \left( \alpha | { \bar{G} }_{ \ocal } , \mathcal{C}, { S }_{ \ocal }^{ \text{prior} } \right) }$ is a probability distribution for $\alpha$ given in Ref.~\cite{jarrell_1996}.\\

 \subsection{The SPA strength function}
 
  The success of the MEM depends on a good choice for the prior distribution.  In this work we explore two choices for the prior strength function: the SPA and the QRPA. In both methods we can calculate directly the strength function without the need for analytic continuation.  The SPA strength function is discussed below (see also in Ref.~\cite{fanto_2024}), while the calculation of the finite-temperature QRPA strength function in the CI shell model framework is discussed in Ref.~\cite{ryssens_2024} and is based on the HF-SHELL code developed in Ref.~\cite{ryssens_2021}.

  The SPA takes into account large-amplitude static fluctuations of the auxiliary fields in the HS transformation and ignore all time-dependent fluctuations.  
 We apply the SPA in the framework of the CI shell model Hamiltonian in which the two-body interaction is written in a separable form~\cite{fanto_2021} 
$\hat H_2 = -(1/2)\sum_\alpha v_\alpha \hat Q_\alpha^2$.  Using the adiabatic approximation of Ref.~\cite{rossignoli_1998}, the SPA strength function of a transition operator $\mathcal{O}_\lambda$  of rank $\lambda$ is given by
\begin{equation} \label{spa_gsf}
S_{\mathcal{O}_\lambda}(T;\omega) \approx \frac{\int d\sigma M(\sigma) Z(\sigma) S_{\mathcal{O}_\lambda}(T,\sigma;\omega)}{\int d\sigma M(\sigma) Z(\sigma)}\,,
\end{equation}
  where $\sigma$ are static auxiliary fields, and $M(\sigma)$ is a measure function given in Ref.~\cite{fanto_2021}. $Z(\sigma)$ is a $\sigma$-dependent partition function,  $Z(\sigma) = \text{Tr}\,e^{-\beta \left(\hat h_\sigma - \sum_{\lambda=p,n} \mu_\lambda \hat N_\lambda\right)}$, where 
  \begin{equation}
  \hat h_\sigma = \hat H_1 - \sum_\alpha v_\alpha \sigma_\alpha \hat Q_\alpha
  \end{equation} 
  is a one-body Hamiltonian, and ${\mu}_{p}$ and ${\mu}_{n}$ are the proton and neutron chemical potentials, respectively. 
  
  $S_{\mathcal{O}_\lambda}(T,\sigma;\omega)$ is the thermal strength function for static fields $\sigma$ given by 
 \begin{equation}\label{sigma_strength}
  \begin{split}
& S_{\mathcal{O}_\lambda}(T,\sigma;\omega) =  S_{\mathcal{O_\lambda}}^{(0)}(T,\sigma)\delta(\omega) \\  &
+ \frac{1}{1- e^{-\beta \omega}} \sum_{\mu, kl}\frac{1}{2} |\mathcal{O}_{\lambda\mu,kl}|^2 (\tilde f_l - \tilde f_k) \delta(\omega -(\tilde E_k - \tilde E_l)) \;,
\end{split}
\end{equation}  
  where $\mathcal{O}_{\lambda\mu,kl} = \langle k | \mathcal{O}_{\lambda \mu} | l \rangle$ are the matrix elements of the one-body operator $\mathcal{O}_{\lambda \mu}$,  $\tilde E_k$ are the generalized quasi-particle energies~\cite{fanto_2021} and $\tilde f_k=(1 + e^{\beta \tilde E_k})^{-1}$ are their corresponding Fermi-Dirac occupation numbers.
  
  The first term on the right-hand-side of Eq.~(\ref{sigma_strength}) is the SPA strength function in the $\omega \to 0$ limit~\cite{rossignoli_1998}
  \begin{equation}\label{strength0}
S_{\mathcal{O_\lambda}}^{(0)}(T,\sigma) = \frac{1}{2}\sum_\mu \sum_{\substack{kl,k^\prime l^\prime\\ E_k = E_l}} \mathcal{O}_{\lambda\mu,kl}^* \mathcal{O}_{\lambda \mu,k^\prime l^\prime} \langle \alpha_k^\dagger \alpha_l \alpha_{k^\prime}^\dagger \alpha_{l^\prime}\rangle_\sigma\;,
\end{equation}
where $\alpha_k, \alpha_k^\dagger$ are, respectively, the annihilation and creation quasiparticles operators in the eigenbasis of $\hat h_\sigma$ with positive energies $E_k$. The expectation values in Eq.~(\ref{strength0}) for a given auxiliary field configuration $\sigma$ are evaluated using Wick's theorem.  We note that the sum in Eq.~(\ref{strength0}) include  both states with $k = l$ as well as distinct quasiparticle states $k\neq l$ that have the same energies $E_k=E_l$.  

The above formulation assumed the grand-canonical ensemble, while the appropriate ensemble to use is the canonical ensemble of fixed number of protons and neutrons.  Rather than implementing a full particle-number projection, we carry out the canonical projection approximately in the saddle-point approximation together with an exact number-parity projection. The latter describes well odd-even effects due to pairing correlations and is carried out using the number-parity projection operator $\hat P_\eta = \Pi_{\nu=p,n}(1+\eta_\nu e^{i \pi \hat N_\nu})/2$, where $\eta_\nu=1$ ($\eta_\nu=-1$) for even (odd) number parity.

In this formulation, the SPA expression for the canonical thermal strength function is
\begin{equation}
    { S }_{ \ocal } (T;\omega) \simeq \frac{ \int_{} d\sigma M(\sigma) \mathcal{ Z }_{ \eta }(\sigma) { S }_{ \ocal,\eta }(T,\sigma;\omega) }{ \int_{} d\sigma M(\sigma) \mathcal{Z }_{ \eta }(\sigma) }\;,
  \end{equation}
  where $\mathcal{Z}_\eta = Z_\eta e^{\sum_{\nu=p,n} (\ln \zeta_\nu -\beta \mu_\nu N_\nu)}$ with
  ${ { Z }_{ \eta }(\sigma) = \text{Tr}\left[ { \hat{ P } }_{ \eta } { e }^{ -\beta \left( { \hat{ h } }_{ \sigma } - \sum_{\nu = \text{p,n}} { \mu }_{ \nu } { \hat{ N } }_{ \nu } \right) } \right] }$ being the number-parity projected one-body partition function and $\zeta_\nu$ saddle-point factors given by Eq.~(25)  of Ref.~\cite{fanto_2021}. The chemical potentials $\mu_p$ and $\mu_n$ are determined by the saddle-point conditions for number of protons $N_p$ and number of neutrons $N_n$, respectively.
  
 $S_{ \ocal,\eta }(T,\sigma;\omega)$, the number-parity projected strength function for a given configuration of the static auxiliary fields $\sigma$, is given by Eqs.~(\ref{sigma_strength}) and (\ref{strength0}) but with the quasi-particle occupation numbers replaced with the number-parity projected occupation numbers $f_k^\eta$ (see Eqs.~(22) and (21) in Ref.~\cite{fanto_2021}), and $\langle \ldots\rangle_\sigma$ replaced by the projected expectation value $\langle \ldots\rangle_{\sigma,\eta}$.

\section{Application to a chain of neodymium isotopes}
\label{section_application_lneodymium}

\subsection{CI shell model space and Hamiltonian}

  We use the theoretical framework of Sec.~\ref{Formalism} to calculate finite-temperature $M1$ strength functions in a chain of even-mass neodymium isotopes ${ {  }^{ 144-152 } }$Nd, which describes the crossover from spherical to well-deformed nuclei. We apply the MEM to extract finite-temperature strength functions from imaginary-time response functions calculated with SMMC, with the aid of prior functions calculated from either SPA or QRPA. 
 
  The CI shell-model single-particle space consists of a set of spherical orbitals ${ \an = ({ n }_{ \an }, { l }_{ \an }, { j }_{ \an }) }$, where ${ \nu = \{{\rm p,n}\} }$.
  The single-particle orbitals and their energies ${ { \epsilon }_{ \an } }$ are taken from a central Woods-Saxon potential plus spin-orbit interaction~\cite{bohr_1969_book}.  The interaction includes a monopole pairing term and multipole-multipole interaction terms. The Hamiltonian is then given by
\begin{align}
  \hat{ H } & = \sum_{ \an } { \epsilon }_{ \an } { \hat{ n } }_{ \an } - \sum_{\nu = p,n} { g }_{ \nu } { \hat{ P } }_{ \nu }^{ \dagger } { \hat{ P } }_{ \nu } \nonumber \\ 
  & - \sum_{ \lambda } { \chi }_{ \lambda } :({ \hat{ O } }_{ \lambda;p } + { \hat{ O } }_{ \lambda;n }).({ \hat{ O } }_{ \lambda;p } + { \hat{ O } }_{ \lambda;n }): 
  \label{}
\end{align}
  with ${ { \hat{ P } }_{ \nu }^{ \dagger } = \sum_{ ({ a }_{ \nu }m) > 0 } { a }_{ \an m }^{ \dagger } { a }_{ \overline{\an m} }^{ \dagger } }$ the ${ J = 0 }$ pair creation operator, where ${ \ket{ \an m } }$ is a shell model single-particle state with magnetic quantum number ${ m }$ for particle species ${ \lambda }$, ${ (\an m) > 0 }$  describes half the single-particle states, and ${ \ket{ \overline{\an m} }= { (-) }^{ { j }_{ \an } + { l }_{ \an } + m } \ket{ \an -m } }$ is the time-reversed state of ${ \ket{ \an m } }$. 
  Here, ${ : : }$ denotes normal ordering and ${ { \hat{ O } }_{ \lambda ; \nu } = \frac{ 1 }{ \sqrt{2\lambda+1} } \sum_{ab} \bra{ { j }_{ \an } } | \frac{ d { V }_{ \text{WS} } }{ dr } { Y }_{ \lambda } | \ket{ { j }_{ \bn } } { \left[ { a }_{ \an }^{ \dagger } \times { \tilde{ a } }_{ \bn } \right] }^{ \lambda } }$ is a surface-peaked multipole operator, with ${ { V }_{ \text{WS} } }$ being the central Woods-Saxon potential.
  We include quadrupole, octupole, and hexadecupole terms (i.e. ${ \lambda = 2,3,4 }$ ) with corresponding strengths ${\chi }_{ \lambda } =  \chi k_ \lambda$. 
  The parameter ${ \chi }$ is determined self-consistently~\cite{alhassid_1996} and ${ { k }_{ \lambda } }$ are renormalization factors accounting for core polarization effects.

  For the lanthanides we use a model space consisting of the following orbitals: ${ 0{ g }_{ 7/2 } }$, ${ 1{ d }_{ 5/2 } }$, ${ 1{ d }_{ 3/2 } }$, ${ 2{ s }_{ 1/2 } }$, ${ 0{ h }_{ 11/2 } }$, and ${ 1{ f }_{ 7/2 } }$ for protons; ${ 0{ h }_{ 11/2 } }$, ${ 0{ h }_{ 9/2 } }$, ${ 1{ f }_{ 7/2 } }$, ${ 1{ f }_{ 5/2 } }$, ${ 2{ p }_{ 3/2 } }$, ${ 2{ p }_{ 1/2 } }$, ${ 0{ i }_{ 13/2 } }$, and ${ 1{ g }_{ 9/2 } }$ for neutrons. 
  The values of the pairing strengths $g_\nu$ and renormalization factors $k_\lambda$ are given in Ref.~\cite{alhassid_2008}.
  
We consider the $M1$ transition operator 
\begin{equation}
  { \hat{ \mathcal{O} } }_{ M1 } = \sqrt{ \frac{ 3 }{ 4 \pi } } \frac{ { \mu }_{ N } }{ \hbar c } ({ g }_{ l } \mathbf{l} + { g }_{ s } \mathbf{s}) \;,
  \label{}
\end{equation}
where ${ \mathbf{l} }$ and ${ \mathbf{s} }$ are the orbital and spin angular momentum operators, respectively.
In our calculations, we used the free-nucleon $g$ factors ${ { g }_{ l,p } = 1}$, ${ { g }_{ l,n } = 0 }$, ${ { g }_{ s,p } = 5.5857 }$, and ${ g = -3.8263 }$.

\subsection{Results} 

\begin{figure*}[bth]
	\includegraphics[width=17.0cm]{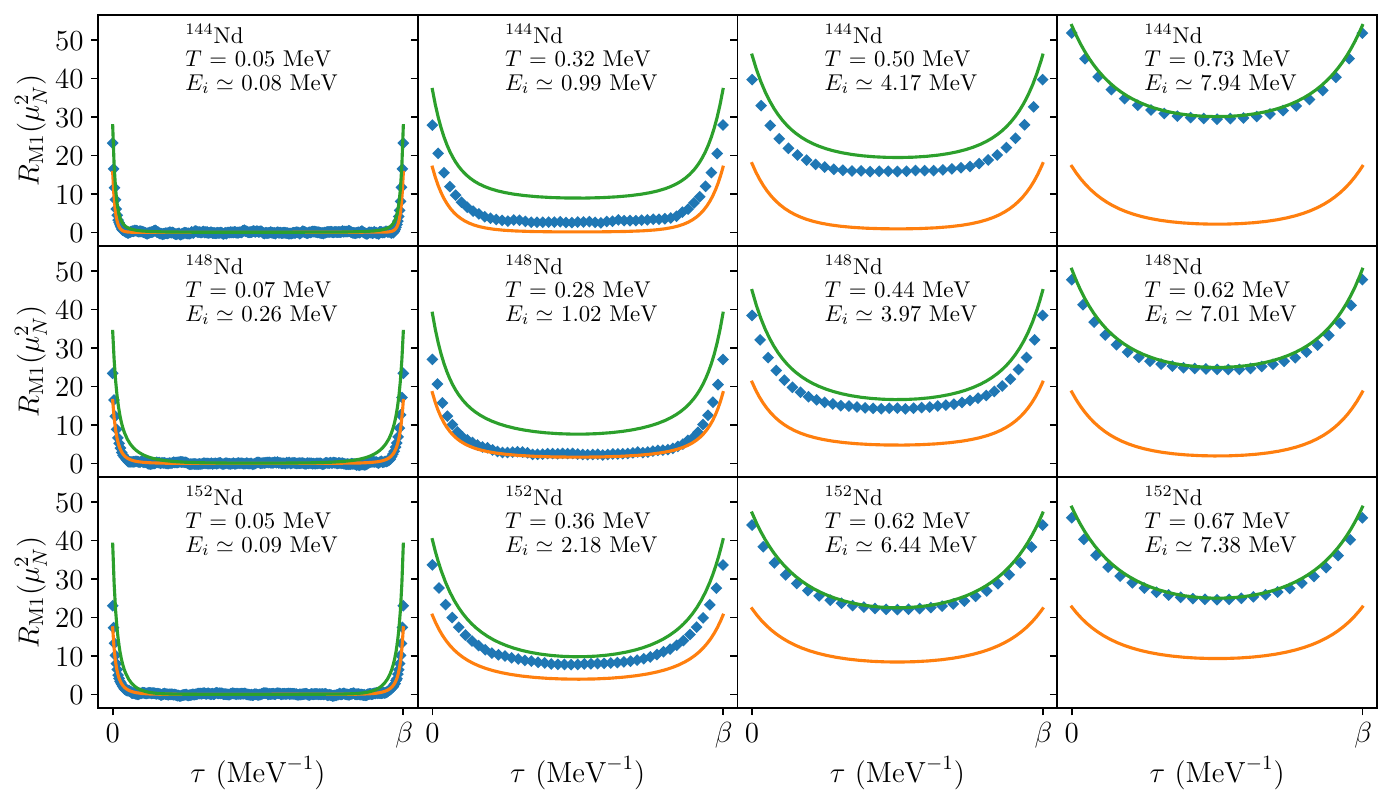} 
        \caption{The $M1$ imaginary-time response function ${R}_{M1}(\tau)$ as a function of imaginary time $\tau$ for a chain of even-mass neodymium isotopes ${}^{144,148,152}$Nd. The left column shows results for near-zero temperatures $T$ (i.e., close to the ground state) while the right column corresponds to the neutron separation energy.  The QRPA (orange solid line) and SPA (green solid line) response functions are compared with the SMMC response functions (blue diamonds). The  SMMC response functions include statistical errors but these errors are smaller than the size of the symbols. }
    	\label{f_response_temperature}
\end{figure*}

We calculated the finite-temperature $M1$ imaginary-time response function $R_{M1}(T;\tau)$ in SMMC using Eq.~(\ref{response-SMMC}). We discretize the imaginary  time interval $[0,\beta]$ into slices of length $\Delta\beta$.  We carried out the calculations for time slices $\Delta \beta=1/32,1/64$ MeV$^{-1}$ and extrapolated to $\Delta \beta=0$.  The MEM is then applied to determine the $M1$ strength function $S_{M1}(T;\tau)$ using for the prior either the SPA or QRPA strength functions. For both priors, we replace the $\delta$ functions [see, e.g., the $\sigma$-dependent SPA strength function in Eq.~(\ref{sigma_strength})] by a Lorentzian with a fixed width parameter of $0.2$ MeV. This width is comparable to the bin width used in conventional CI shell-model studies~\cite{schwengner_2013,karampagia_2017,schwengner_2017,sieja_2017,mitdbo_2018}.   In the following, we refer to the MEM with the QRPA prior as MEM-QRPA, and the MEM with the SPA prior as MEM-SPA. 

\subsubsection{$M1$ response and strength functions using the QRPA and SPA priors} 

The reliability of MEM depends upon a good choice of the prior strength function.
In Fig.~\ref{f_response_temperature} we compare the SMMC $M1$ imaginary-time response function $R_{M1}$ (blue solid diamonds) with the SPA (green lines) and QRPA (orange lines) response functions at several temperatures for spherical ($^{144}$Nd), transitional  ($^{148}$Nd) and deformed  ($^{152}$Nd) neodymium isotopes.  $E_i$ in the panels denotes the initial excitation energy of the nucleus which in the finite-temperature formalism is given by the average thermal energy $E(T)$ (measured with respect to the ground-state energy).   We observe that at very low temperatures close to the ground state (left column in Fig.~\ref{f_response_temperature} both the QRPA and SPA response functions are quite close to the exact SMMC results, with the QRPA response function being essentially on top of the SMMC response function. 
However, with increasing temperature, the SPA response function is in closer agreement with the SMMC results, when compared to the QRPA response functions.  In particular at temperatures that correspond to the neutron separation energy of $E_i  \sim 7 -8$ MeV (right column in Fig.~\ref{f_response_temperature}), the SPA is a superior approximation to the QRPA.  We conclude that large-amplitude static fluctuations around the mean field that are accounted for in the SPA but are missing in the QRPA play an important role at finite temperature.

 \begin{figure*}[bth]
	\includegraphics[width=17.0cm]{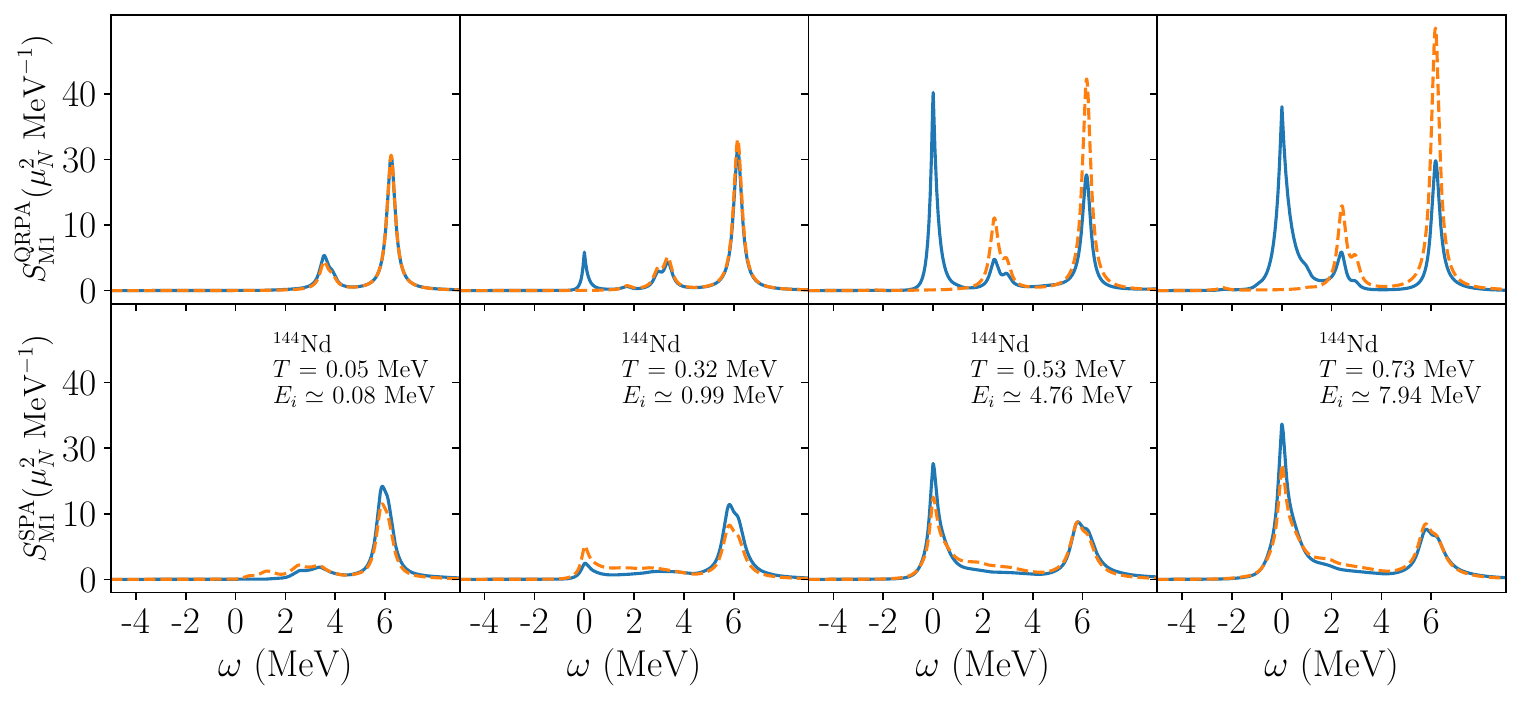} 
	\caption{Strength function $S_{M1}(\omega)$ for $^{ 144 }$Nd at different temperatures, from close to the ground state ($T=0.05$ MeV) up to a temperature of  $T=0.73$ MeV, which corresponds to the neutron separation energy. The MEM results (blue solid lines) are compared with the prior results (orange dashed lines). The top row shows results with the QRPA strength as prior, while the bottom row shows results with the SPA prior.}
 \label{f_Nd144_temps}
\end{figure*}

In Fig.~\ref{f_Nd144_temps} we show the temperature dependence of the $M1$ strength function $S_{M1}$ versus transition energy $\omega$ for the spherical isotope $^{144}$Nd.  The MEM strength functions are shown in blue solid lines while the prior strength functions are the orange dashed lines.  In the top row the prior is the QRPA strength function while in the bottom row the prior is the SPA strength function. The various temperatures correspond  to average initial excitation energies $E_i$ shown in the panels.  The near ground-state strength functions ($T = 0.05$ MeV) obtained with QRPA and SPA are similar.  For the QRPA, the MEM strength function essentially coincides with the QRPA prior, while for the SPA we observe a small difference.  As the temperature increases towards the neutron separation energy $\sim 7.94$ MeV, we observe larger differences between the MEM and prior strength functions in the QRPA case as compared with the SPA case.  This is consistent with the comparison between the various response functions shown in Fig.~\ref{f_response_temperature}. We conclude that the strength function at very low temperatures near the ground state is well described by the QRPA while the SPA is a better prior to use at the neutron resonance energy.

As the initial excitation energy increases, we notice the appearance of an $\omega=0$ peak that becomes larger as we approach the neutron resonance energy.  This will be interpreted as the LEE. 

In general the response function is less sensitive to the behavior of the strength function at large values of $\omega$. 
This is attributed to the exponential suppression of the large-$\omega$  contribution in the Laplace transform. 
On the other hand, we expect the low-$\omega$ results and in particular the LEE to be highly reliable. We note that the MEM strength function reproduces the LEE even for the QRPA prior that do not exhibits a LEE.

\subsubsection{Ground-state $M1$ strength functions}

The ground-state $M1$ strength functions for the chain of even-mass neodymium isotopes $^{144-152}$Nd are shown in Fig.~\ref{f_isotopes_GS} using very low temperatures.  The top row corresponds to the QRPA prior and the bottom row corresponds to the SPA prior.  For each prior we show the corresponding strength functions by orange dashed lines, while the MEM strength functions are described by the blue solid lines. The overall structure of the MEM strength functions exhibits similar behavior when using either the QRPA or the SPA strength functions as prior. The deviation of the MEM from the prior is smaller for the QRPA, confirming our earlier conclusion that at very low temperatures the QRPA is a better approximation than the SPA. 

In general the strength function $S_{M1}$ describes the absorption of $\gamma$-rays for $\omega >0$ and their emission for $\omega <0$. For transitions from the ground state the strength is non-zero for only $\omega >0$.   

For the spherical isotopes $^{144}$Nd and $^{146}$Nd we observe in Fig.~\ref{f_isotopes_GS} a pronounced peak at $\sim 6$ MeV, which we interpret as the spin-flip mode.  This mode involves a transitions of the type $(j=l+\half) \to (j = l - \half)$ and is characterized by the flip of one or more nuclear spins, which results in a change in the total angular momentum of the nucleus. The strength of the spin-flip mode becomes fragmented in the transitional $^{148}$Nd isotope and in the deformed $^{150-152}$Nd isotopes. 

We also observe in the ground-state $M1$ strength function a structure around $\omega \simeq 2-3$ MeV, which we interpret as the scissors mode. This mode describes the collective oscillations of the proton and neutron clouds against each other like scissors blades~\cite{heyde_2010}.

\begin{figure*}[bth]
	\includegraphics[width=17.0cm]{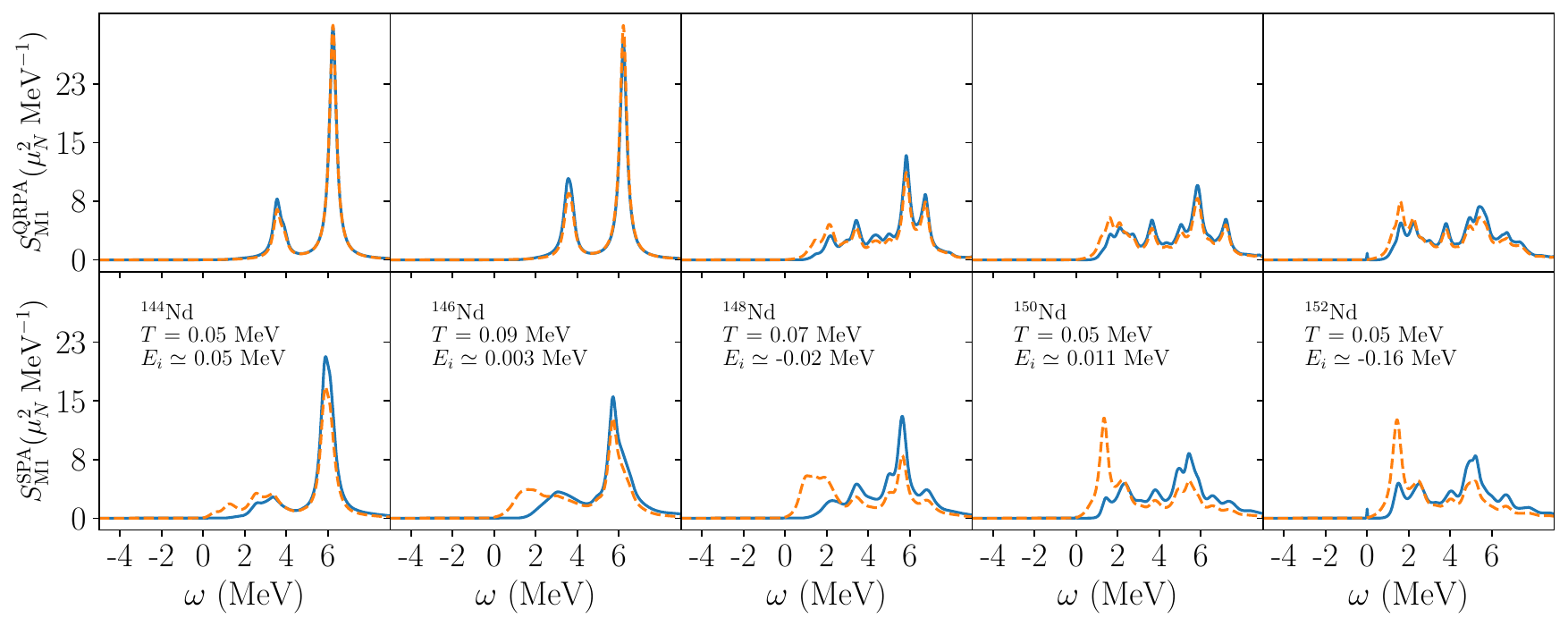} 
	\caption{Near ground-state strength functions $S_{\rm M1}(\omega)$ versus $\omega$ for the even-mass $^{ 144-152}$Nd isotopes. The MEM results (blue solid lines) are compared with the prior results (orange dashed lines). The top row shows results with the  QRPA  prior, while the bottom row shows results with the SPA prior.}
    	\label{f_isotopes_GS}
\end{figure*}

\subsubsection{Finite-temperature $M1$  strength functions}

As the temperature $T$ increases (which translates into increasing initial excitation energy $E_i$ of the nucleus), a peak centered at $\omega=0$ develops in the strength function. This is clearly seen in Fig.~\ref{f_Nd144_temps}) for $^{144}$Nd. We see such a peak already $T=0.32$ MeV (corresponding to $E_i \simeq 0.96$ MeV) and it becomes more pronounced with increasing temperature. We will identify this peak with the LEE observed in the deexcitation strength function at low $\gamma$-ray energies $E_\gamma$.  We note that the SPA strength function already reproduces the LEE and the MEM-SPA makes it more enhanced. On the other hand, the LEE is missing in the QRPA strength function but appears in the MEM-QRPA. This suggests that the LEE is a very robust feature of the finite-temperature $M1$ strength function. 

In Fig.~\ref{f_isotopes_NSE}, we show the $M1$ strength functions for the even-mass neodymium isotopes for initial energies that correspond to the neutron separation energy. The orange dashed lines are the SPA strength functions, while the blue solid lines are the MEM-SPA.  We limit our calculations to the SPA prior since at the neutron resonance energy, the SPA response function is  closer to the SMMC response function when compared with the QRPA response function  (see Fig~\ref{f_response_temperature}). We observe a LEE structure for all the neodymium isotopes with the LEE peak decreasing with increased neutron number, i.e., in the crossover from spherical to deformed nuclei.  A slight shallow structure, which we interpret as the scissors mode built on top of excited states, seems to develop close to $\sim 2$ MeV in the deformed isotopes.  We conclude that some of the LEE strength is transferred to the scissors mode as the nucleus becomes more deformed. This  scissors mode is more noticeable in the finite-temperature QRPA calculations as is seen in the QRPA results for the spherical nucleus $^{144}$Nd in Fig.~\ref{f_Nd144_temps}).  A scissors mode built on top of excited states has been observed experimentally in lanthanide nuclei~\cite{krticka_2011}.  The increase in the scissors mode in deformed nuclei has been observed experimentally at the neutron separation energy  in neodymium isotopes~\cite{guttormsen_2022},

A spin-flip mode at $\sim 6$ MeV is observed in all of the neodymium isotopes also at the neutron separation energy (see Fig.~\ref{f_isotopes_NSE}) and its height decreases in the crossover from spherical to deformed neodymium isotopes.

\begin{figure*}[bth]
	\includegraphics[width=17.0cm]{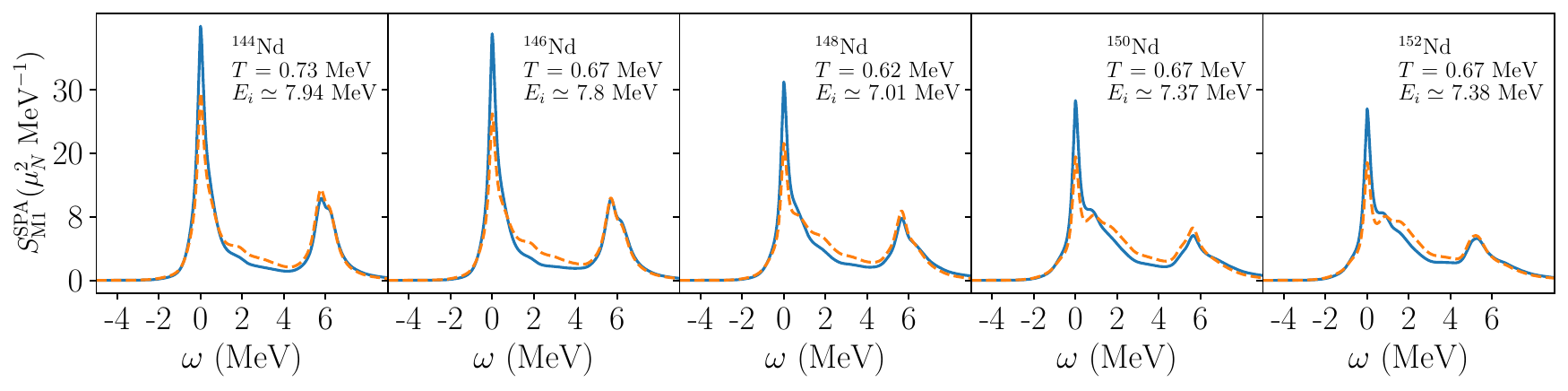} 
	\caption{Strength functions $S_{\rm M1}(\omega)$ versus $\omega$ for the even-mass $^{ 144-152}$Nd isotopes at temperatures that correspond to the neutron separation energies. The orange dashed lines are the SPA strength functions and the blue solid lines are the MEM strength functions using the SPA as prior.}
    	\label{f_isotopes_NSE}
\end{figure*}

 \begin{figure*}[bth]
	\includegraphics[width=17.0cm]{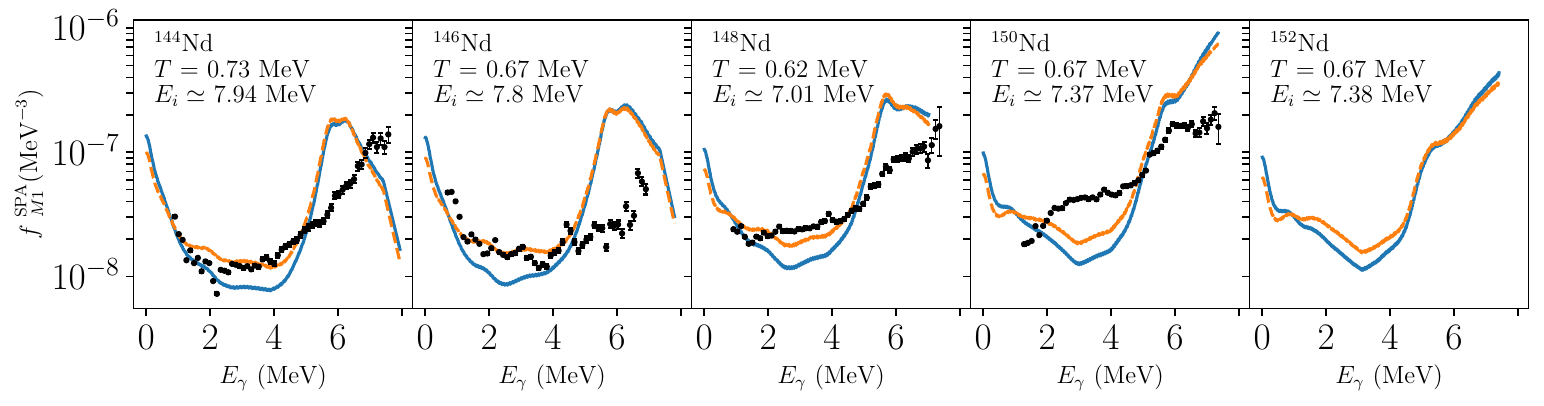} 
	\caption{Deexcitation ${ \gamma }$-ray strength functions $f_{\rm M1}$ as a function of the $\gamma$-ray energy $E_\gamma$ at the neutron separation energy for the even-mass neodymium isotopes $^{144-152}$Nd. The MEM results (blue solid lines) are compared with the SPA prior results (orange dashed lines) and with the experimental data (black symbols) of Ref.~\cite{guttormsen_2022}.}
    	\label{f_GSF_NSE}
\end{figure*}

\subsubsection{Deexcitation $M1$ strength function $f_{\rm M1}$ and the LEE}

In experiments, one usually measures the deexcitation $M1$ strength function, $f_{\rm M1}$.  This function describes the probability for the compound nucleus to decay by $M1$ $\gamma$-rays with energy $E_\gamma$. The deexcitation strength $f_{\rm M1}$ for initial energy $E_i$ and $\gamma$-ray energy $E_\gamma$  is related to the strength function $S_{\rm M1}$ by~\cite{fanto_2024}
\begin{equation}\label{f_M1}
f_{M1}(E_i, E_\gamma)  \approx \frac{1}{3} a \frac{\tilde \rho (E_i)}{\tilde \rho(E_i -E_\gamma)} S_{M1}(T; \omega=-E_\gamma)\,,
\end{equation}
where  $a= \frac{16 \pi}{9(\hbar c)^3}$, $T$ is the temperature that corresponds to an initial excitation energy $E_i$, and $\tilde \rho(E_x)$ is the level density at excitation energy $E_x$. The factor of $1/3$ takes into account the 3 possible value of the final spin for given initial value of the spin~\cite{dallas}.

Using SMMC level densities~\cite{guttormsen_2021} and the MEM strength functions $S_{M1}(T;\omega)$, we calculated the deexcitation strength function $f_{\rm M1}$ using Eq.~(\ref{f_M1}) for the even mass ${}^{144-152}$Nd at an initial energy that corresponds to the neutron separation energy. In Fig.~\ref{f_GSF_NSE} we show $f_{\rm M1}$ versus $E_\gamma$. The dashed orange lines are the SPA $f_{\rm M1}$ and the blue solid lines are the MEM-SPA results.  We clearly observed the LEE at low $\gamma$-ray energies (below $\sim 2$ MeV) for all the even-mass neodymium isotopes. 

We compare our results with recent experimental data for the deexcitation $\gamma$SF~\cite{guttormsen_2022}. We note, however, that the experimental results include contributions from both $E1$ and $M1$ and a detailed comparison with experiment requires the calculation of the $E1$   $\gamma$SF. 
 In the spherical isotopes ${}^{144,146}$Nd, a clear LEE is observed experimentally, in overall agreement with our theoretical results.  In the transitional isotope $^{148}$Nd, a LEE seems to develop below $\sim 1.5$ MeV. No LEE is observed experimentally in the deformed isotope $^{150}$Nd while our calculated $f_{M1}$ clearly indicates the existence of a LEE. We note, however, the absence of experimental data at low $\gamma$-ray energies. Thus the missing LEE in the experimental results for $^{150}$Nd does not preclude the possibility of a residual LEE in neutron-rich neodymium isotopes. In fact, our calculations suggest its persistence in ${}^{152}$Nd.
 
 \begin{figure}[bth]
	\includegraphics[width=8.0cm]{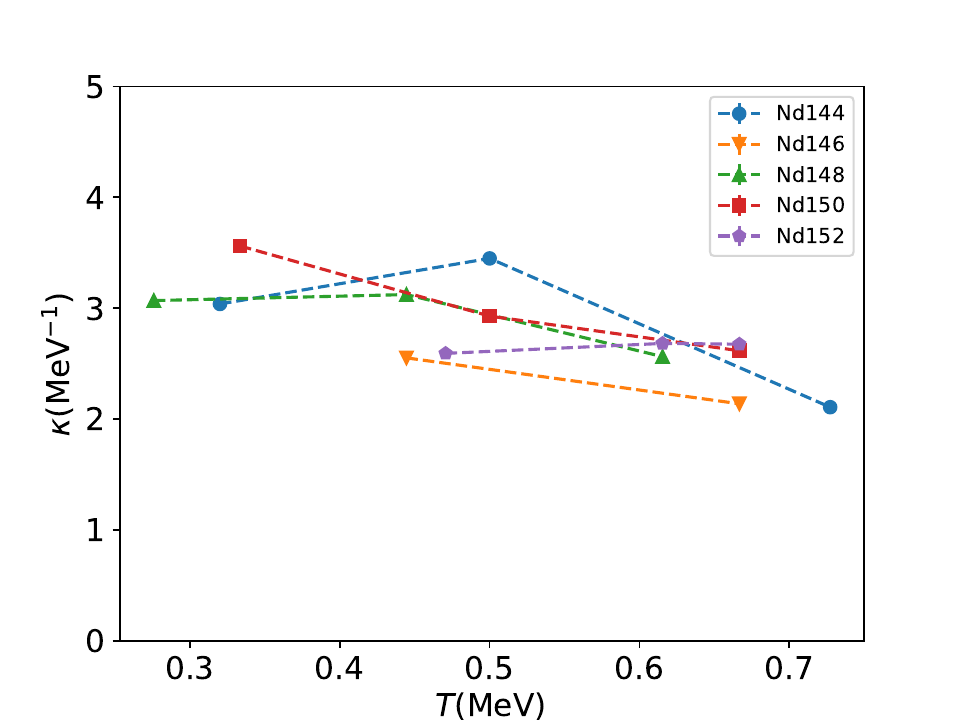} 
	\caption{The LEE slope $\kappa$ (on a logarithmic scale) as a function of temperature $T$ for the even-mass neodymium isotopes $^{144-152}$Nd in the MEM-SPA calculations.}
    	\label{f_slope_as_temp}
\end{figure}

As suggested in Refs.~\cite{schwengner_2013,karampagia_2017}, we find that the LEE is well described by an exponential form~\cite{schwengner_2013} 
\begin{equation}
f_{\rm M1}(E_\gamma) \approx C_0 {e}^{-\kappa {E}_{\gamma}} \;,
\end{equation}
 where $C_0$ is the value of the LEE at $E_\gamma=0$ and  $\kappa$ describes the slope of the LEE on a logarithmic scale.  
In Fig.~\ref{f_slope_as_temp}, the values of $\kappa$ for ${}^{144-152}$Nd are plotted as a function of temperature $T$.
The results show that $\kappa$ remains roughly constant across a wide range of temperatures for each the neodymium isotopes. 
This finding, which is consistent with conventional CI shell-model results in smaller model space~\cite{karampagia_2017}, indicates that the slope of the LEE depends only weakly on the initial energy.  On average, the $\kappa$ values for the different isotopes are similar.

Strictly speaking, in the micorcanonical ensemble for which the initial energy is fixed, the temperature fluctuates and its variance is given by $\langle(\Delta T)^2\rangle = T^2 /C$, where $C$ is the canonical heat capacity. For the neodymium isotopes the r.m.s. $\sqrt{\langle(\Delta T)^2\rangle}$ of temperature fluctuations varies between $0$ at $T=0$ and $\sim 0.15 - 0.18$ MeV at the neutron separation energies. Since the LEE depends only weakly on temperature, we expect our results for the LEE  to be robust despite these temperature fluctuations.  

In Table~\ref{tab_slope_GSF} we list the values of $\kappa$ at the neutron separation energy. We find that the slope $\kappa$ increases while the LEE maximal value $C_0$ decreases with the onset of deformation along the chain of neodymium  isotopes. 
In particular, we find close agreement between our value of $\kappa =2.1$ MeV$^{-1}$ in the spherical isotopes ${}^{144,146}$Nd and the experimental value of $\kappa = 1.9$ MeV$^{-1}$~\cite{guttormsen_2022}.\\

\begin{table}[h]
    \centering
    \begin{tabular}{ |c|c|c| }
    \hline
     & $\kappa$ (${\text{MeV}}^{-1}$) & $C_0$ (${\text{MeV}}^{-3}$) \\ \hline
    & & \\ [-0.5em]
    ${}^{144}$Nd & 2.1 & $4.1\times{10}^{-7}$ \\
    ${}^{146}$Nd & 2.1 & $3.9\times{10}^{-7}$ \\
    ${}^{148}$Nd & 2.6 & $3.3\times{10}^{-7}$ \\
    ${}^{150}$Nd & 2.6 & $3.1\times{10}^{-7}$ \\
    ${}^{152}$Nd & 2.7 & $2.9\times{10}^{-7}$ \\
    \hline
\end{tabular}
    \caption{LEE slopes $\kappa$ and LEE maximal values $C_0$ at $E_\gamma=0$ obtained from fits to the deexcitation strength function $f_{\rm M1}$ in Fig.~\ref{f_GSF_NSE}  at low $\gamma$-ray energies.}
    \label{tab_slope_GSF}
\end{table}

\begin{comment}
    \begin{table}[]
    \centering
    \begin{tabular}{ |c|c|c| } 
    \hline 
    $T$ (MeV) & ${\kappa}_{\text{QRPA}}$ (${\text{MeV}}^{-1}$) & ${\kappa}_{\text{SPA}}$ (${\text{MeV}}^{-1}$) \\
    \hline
    0.05 & 9.8 & 8.1 \\ 
    0.32 & 5.6 & 0.9 \\ 
    0.53 & 4.6 & 2.0 \\ 
    0.73 & 2.2 & 1.8 \\ 
    \hline
\end{tabular}
    \caption{LEE slopes from Fig.(\ref{f_Nd144_temps}).}
    \label{tab:my_label}
\end{table}

\begin{table}[]
    \centering
    \begin{tabular}{ |c|c|c| } 
    \hline 
     & ${\kappa}_{\text{QRPA}}$ (${\text{MeV}}^{-1}$) & ${\kappa}_{\text{SPA}}$ (${\text{MeV}}^{-1}$) \\
    \hline
    ${}^{144}$Nd & 2.2 & 1.8 \\ 
    ${}^{146}$Nd & 3.2 & 1.8 \\ 
    ${}^{148}$Nd & 1.1 & 1.2 \\ 
    ${}^{150}$Nd & 0.4 & 0.8 \\
    ${}^{152}$Nd & 0.4 & 0.7 \\
    \hline
\end{tabular}
    \caption{LEE slopes from Fig.(\ref{f_isotopes_NSE}).}
    \label{tab:my_label}
\end{table}
\end{comment}

\section{Conclusions}
\label{concl}

In this study, we employed the MEM to deduce $M1$ $\gamma$SF in the even-mass $^{144-152}$Nd isotopes from the exact SMMC imaginary-time $M1$ response functions, using the SPA and QRPA strength functions as prior. We find that at low temperatures (close to the ground state) the QRPA response function is somewhat closer to the SMMC response  function than the corresponding SPA, but that at higher temperatures and, in particular close to the neutron resonance energy, the SPA is a better starting point to use as prior.   We identify a spin-flip mode in both the ground state and finite-temperature $M1$ $\gamma$SF. A scissors mode, clearly visible in the ground-state $\gamma$SF, is suppressed at finite temperature but is still visible in the deformed nuclei.  

At finite temperature, and, in particular, at the neutron separation energy, a LEE structure develops in the $\gamma$SF of all the even-mass neodymium isotopes ${ {}^{144-152} }$Nd. The peak of the LEE is strongest in the spherical isotopes and decreases in the deformed nuclei  with some of its strength transferring to the scissors mode at $\omega \approx 2-3$ MeV built on top of excited states. The LEE structure is not reproduced in the finite-temperature QRPA strength function but is observed in the SPA strength function. This LEE structure is very robust and emerges also in the MEM with the QRPA prior despite its absence in the QRPA strength function.
In addition to the LEE and the scissors mode, we observe a spin-flip mode around $\omega \approx 6$ MeV, with its strength decreasing with deformation. 

We also calculated the deexcitation strength function $f_{M1}$ as a function of the $\gamma$-ray energy $E_\gamma$ at the neutron separation energy, and identified a LEE structure in all the neodymium isotopes under study. Our results are overall consistent with the experimental findings of Ref.~\cite{guttormsen_2022}. The latter did not observe a LEE in the deformed isotope $^{150}$Nd but this might be due to the lack of data at the very low $\gamma$-ray energies. We identify a LEE also in the deformed $^{152}$Nd for which there are no experimental data.  We also find that the slope of the LEE (on a logarithmic scale) is approximately independent of the initial temperature (or excitation energy). 

Currently our method is the only one capable of reproducing a LEE in heavy open-shell nuclei for which conventional CI shell-model calculations are prohibited.  If the LEE persists in heavy neutron-rich nuclei, it will have significant effects on $r$-process nucleosynthesis by enhancing the neutron-capture rates. 

\begin{acknowledgments}

This work was supported in part by the U.S. DOE grant No.~DE-SC0019521.
The calculations used resources of the National Energy Research Scientific Computing Center (NERSC), a U.S. Department of Energy Office of Science User Facility operated under Contract No.~DE-AC02-05CH11231.  We thank the Yale Center for Research Computing for guidance and use of the research computing infrastructure. W. R. is a Research Associate of the F.R.S.-FNRS and a member of BLU-ULB (Brussels laboratory of the Universe).

\end{acknowledgments}

%

%\bibliographystyle{apsrev4-1}
%\bibliography{refs}

\begin{thebibliography}{63}%
\makeatletter
\providecommand \@ifxundefined [1]{%
 \@ifx{#1\undefined}
}%
\providecommand \@ifnum [1]{%
 \ifnum #1\expandafter \@firstoftwo
 \else \expandafter \@secondoftwo
 \fi
}%
\providecommand \@ifx [1]{%
 \ifx #1\expandafter \@firstoftwo
 \else \expandafter \@secondoftwo
 \fi
}%
\providecommand \natexlab [1]{#1}%
\providecommand \enquote  [1]{``#1''}%
\providecommand \bibnamefont  [1]{#1}%
\providecommand \bibfnamefont [1]{#1}%
\providecommand \citenamefont [1]{#1}%
\providecommand \href@noop [0]{\@secondoftwo}%
\providecommand \href [0]{\begingroup \@sanitize@url \@href}%
\providecommand \@href[1]{\@@startlink{#1}\@@href}%
\providecommand \@@href[1]{\endgroup#1\@@endlink}%
\providecommand \@sanitize@url [0]{\catcode `\\12\catcode `\$12\catcode
  `\&12\catcode `\#12\catcode `\^12\catcode `\_12\catcode `\%12\relax}%
\providecommand \@@startlink[1]{}%
\providecommand \@@endlink[0]{}%
\providecommand \url  [0]{\begingroup\@sanitize@url \@url }%
\providecommand \@url [1]{\endgroup\@href {#1}{\urlprefix }}%
\providecommand \urlprefix  [0]{URL }%
\providecommand \Eprint [0]{\href }%
\providecommand \doibase [0]{http://dx.doi.org/}%
\providecommand \selectlanguage [0]{\@gobble}%
\providecommand \bibinfo  [0]{\@secondoftwo}%
\providecommand \bibfield  [0]{\@secondoftwo}%
\providecommand \translation [1]{[#1]}%
\providecommand \BibitemOpen [0]{}%
\providecommand \bibitemStop [0]{}%
\providecommand \bibitemNoStop [0]{.\EOS\space}%
\providecommand \EOS [0]{\spacefactor3000\relax}%
\providecommand \BibitemShut  [1]{\csname bibitem#1\endcsname}%
\let\auto@bib@innerbib\@empty
%</preamble>
\bibitem [{\citenamefont {Hauser}\ and\ \citenamefont
  {Feshbach}(1952)}]{hauser_1952}%
  \BibitemOpen
  \bibfield  {author} {\bibinfo {author} {\bibfnamefont {W.}~\bibnamefont
  {Hauser}}\ and\ \bibinfo {author} {\bibfnamefont {H.}~\bibnamefont
  {Feshbach}},\ }\href {\doibase 10.1103/PhysRev.87.366} {\bibfield  {journal}
  {\bibinfo  {journal} {Phys. Rev.}\ }\textbf {\bibinfo {volume} {87}},\
  \bibinfo {pages} {366} (\bibinfo {year} {1952})}\BibitemShut {NoStop}%
\bibitem [{\citenamefont {Koning}\ and\ \citenamefont
  {Rochman}(2012)}]{koonin_2012}%
  \BibitemOpen
  \bibfield  {author} {\bibinfo {author} {\bibfnamefont {A.~J.}\ \bibnamefont
  {Koning}}\ and\ \bibinfo {author} {\bibfnamefont {D.}~\bibnamefont
  {Rochman}},\ }\href@noop {} {\bibfield  {journal} {\bibinfo  {journal} {Nucl.
  Data Sheets}\ }\textbf {\bibinfo {volume} {113}},\ \bibinfo {pages} {2841}
  (\bibinfo {year} {2012})}\BibitemShut {NoStop}%
\bibitem [{\citenamefont {Bartholomew}\ \emph {et~al.}()\citenamefont
  {Bartholomew}, \citenamefont {Earle}, \citenamefont {Ferguson}, \citenamefont
  {Knowles},\ and\ \citenamefont {Lone}}]{bartholomev_1973}%
  \BibitemOpen
  \bibfield  {author} {\bibinfo {author} {\bibfnamefont {G.~A.}\ \bibnamefont
  {Bartholomew}}, \bibinfo {author} {\bibfnamefont {E.~D.}\ \bibnamefont
  {Earle}}, \bibinfo {author} {\bibfnamefont {A.~J.}\ \bibnamefont {Ferguson}},
  \bibinfo {author} {\bibfnamefont {J.~W.}\ \bibnamefont {Knowles}}, \ and\
  \bibinfo {author} {\bibfnamefont {M.~A.}\ \bibnamefont {Lone}},\ }\href@noop
  {} {\emph {\bibinfo {title} {Advances in Nuclear Physics}}},\ edited by\
  \bibinfo {editor} {\bibfnamefont {M.}~\bibnamefont {Baranger}}\ and\ \bibinfo
  {editor} {\bibfnamefont {E.}~\bibnamefont {Vogt}}\ (\bibinfo  {publisher}
  {Springer Boston 1973})\ pp.\ \bibinfo {pages} {267--298}\BibitemShut
  {NoStop}%
\bibitem [{\citenamefont {Mumpower}\ \emph {et~al.}(2016)\citenamefont
  {Mumpower}, \citenamefont {Surman}, \citenamefont {McLaughlin},\ and\
  \citenamefont {Aprahamian}}]{mumpower_2016}%
  \BibitemOpen
  \bibfield  {author} {\bibinfo {author} {\bibfnamefont {M.}~\bibnamefont
  {Mumpower}}, \bibinfo {author} {\bibfnamefont {R.}~\bibnamefont {Surman}},
  \bibinfo {author} {\bibfnamefont {G.~C.}\ \bibnamefont {McLaughlin}}, \ and\
  \bibinfo {author} {\bibfnamefont {A.}~\bibnamefont {Aprahamian}},\
  }\href@noop {} {\bibfield  {journal} {\bibinfo  {journal} {Prog. Part. Nucl.
  Phys.}\ }\textbf {\bibinfo {volume} {86}},\ \bibinfo {pages} {86} (\bibinfo
  {year} {2016})}\BibitemShut {NoStop}%
\bibitem [{\citenamefont {Bohle}\ \emph {et~al.}(1984)\citenamefont {Bohle},
  \citenamefont {Richter}, \citenamefont {Steffen}, \citenamefont {Dieperink},
  \citenamefont {Iudice}, \citenamefont {Palumbo},\ and\ \citenamefont
  {Scholten}}]{bohle_1984}%
  \BibitemOpen
  \bibfield  {author} {\bibinfo {author} {\bibfnamefont {D.}~\bibnamefont
  {Bohle}}, \bibinfo {author} {\bibfnamefont {A.}~\bibnamefont {Richter}},
  \bibinfo {author} {\bibfnamefont {W.}~\bibnamefont {Steffen}}, \bibinfo
  {author} {\bibfnamefont {A.~E.~L.}\ \bibnamefont {Dieperink}}, \bibinfo
  {author} {\bibfnamefont {N.~L.}\ \bibnamefont {Iudice}}, \bibinfo {author}
  {\bibfnamefont {F.}~\bibnamefont {Palumbo}}, \ and\ \bibinfo {author}
  {\bibfnamefont {O.}~\bibnamefont {Scholten}},\ }\href@noop {} {\bibfield
  {journal} {\bibinfo  {journal} {Phys. Lett. B}\ }\textbf {\bibinfo {volume}
  {137}},\ \bibinfo {pages} {27} (\bibinfo {year} {1984})}\BibitemShut
  {NoStop}%
\bibitem [{\citenamefont {Heyde}\ \emph {et~al.}(2010)\citenamefont {Heyde},
  \citenamefont {von Neumann-Cosel},\ and\ \citenamefont
  {Richter}}]{heyde_2010}%
  \BibitemOpen
  \bibfield  {author} {\bibinfo {author} {\bibfnamefont {K.}~\bibnamefont
  {Heyde}}, \bibinfo {author} {\bibfnamefont {P.}~\bibnamefont {von
  Neumann-Cosel}}, \ and\ \bibinfo {author} {\bibfnamefont {A.}~\bibnamefont
  {Richter}},\ }\href@noop {} {\bibfield  {journal} {\bibinfo  {journal} {Rev.
  Mod. Phys.}\ }\textbf {\bibinfo {volume} {82}},\ \bibinfo {pages} {2365}
  (\bibinfo {year} {2010})}\BibitemShut {NoStop}%
\bibitem [{\citenamefont {Mumpower}\ \emph {et~al.}(2017)\citenamefont
  {Mumpower}, \citenamefont {Kawano}, \citenamefont {Ullmann}, \citenamefont
  {Krti\ifmmode~\check{c}\else \v{c}\fi{}ka},\ and\ \citenamefont
  {Sprouse}}]{mumpower_2017}%
  \BibitemOpen
  \bibfield  {author} {\bibinfo {author} {\bibfnamefont {M.~R.}\ \bibnamefont
  {Mumpower}}, \bibinfo {author} {\bibfnamefont {T.}~\bibnamefont {Kawano}},
  \bibinfo {author} {\bibfnamefont {J.~L.}\ \bibnamefont {Ullmann}}, \bibinfo
  {author} {\bibfnamefont {M.}~\bibnamefont {Krti\ifmmode~\check{c}\else
  \v{c}\fi{}ka}}, \ and\ \bibinfo {author} {\bibfnamefont {T.~M.}\ \bibnamefont
  {Sprouse}},\ }\href {\doibase 10.1103/PhysRevC.96.024612} {\bibfield
  {journal} {\bibinfo  {journal} {Phys. Rev. C}\ }\textbf {\bibinfo {volume}
  {96}},\ \bibinfo {pages} {024612} (\bibinfo {year} {2017})}\BibitemShut
  {NoStop}%
\bibitem [{\citenamefont {Voinov}\ \emph {et~al.}(2004)\citenamefont {Voinov},
  \citenamefont {Algin}, \citenamefont {Agvaanluvsan}, \citenamefont {Belgya},
  \citenamefont {Chankova}, \citenamefont {Guttormsen}, \citenamefont
  {Mitchell}, \citenamefont {Rekstad}, \citenamefont {Schiller},\ and\
  \citenamefont {Siem}}]{voinov_2004}%
  \BibitemOpen
  \bibfield  {author} {\bibinfo {author} {\bibfnamefont {A.}~\bibnamefont
  {Voinov}}, \bibinfo {author} {\bibfnamefont {E.}~\bibnamefont {Algin}},
  \bibinfo {author} {\bibfnamefont {U.}~\bibnamefont {Agvaanluvsan}}, \bibinfo
  {author} {\bibfnamefont {T.}~\bibnamefont {Belgya}}, \bibinfo {author}
  {\bibfnamefont {R.}~\bibnamefont {Chankova}}, \bibinfo {author}
  {\bibfnamefont {M.}~\bibnamefont {Guttormsen}}, \bibinfo {author}
  {\bibfnamefont {G.~E.}\ \bibnamefont {Mitchell}}, \bibinfo {author}
  {\bibfnamefont {J.}~\bibnamefont {Rekstad}}, \bibinfo {author} {\bibfnamefont
  {A.}~\bibnamefont {Schiller}}, \ and\ \bibinfo {author} {\bibfnamefont
  {S.}~\bibnamefont {Siem}},\ }\href@noop {} {\bibfield  {journal} {\bibinfo
  {journal} {Phys. Rev. Lett.}\ }\textbf {\bibinfo {volume} {93}},\ \bibinfo
  {pages} {142504} (\bibinfo {year} {2004})}\BibitemShut {NoStop}%
\bibitem [{\citenamefont {Guttormsen}\ \emph {et~al.}(2005)\citenamefont
  {Guttormsen}, \citenamefont {Chankova}, \citenamefont {Agvaanluvsan},
  \citenamefont {Algin}, \citenamefont {Bernstein}, \citenamefont
  {Ingebretsen}, \citenamefont {L\"onnroth}, \citenamefont {Messelt},
  \citenamefont {Mitchell}, \citenamefont {Rekstad}, \citenamefont {Schiller},
  \citenamefont {Siem}, \citenamefont {Sunde}, \citenamefont {Voinov},\ and\
  \citenamefont {\O{}deg\aa{}rd}}]{guttormsen_2005}%
  \BibitemOpen
  \bibfield  {author} {\bibinfo {author} {\bibfnamefont {M.}~\bibnamefont
  {Guttormsen}}, \bibinfo {author} {\bibfnamefont {R.}~\bibnamefont
  {Chankova}}, \bibinfo {author} {\bibfnamefont {U.}~\bibnamefont
  {Agvaanluvsan}}, \bibinfo {author} {\bibfnamefont {E.}~\bibnamefont {Algin}},
  \bibinfo {author} {\bibfnamefont {L.~A.}\ \bibnamefont {Bernstein}}, \bibinfo
  {author} {\bibfnamefont {F.}~\bibnamefont {Ingebretsen}}, \bibinfo {author}
  {\bibfnamefont {T.}~\bibnamefont {L\"onnroth}}, \bibinfo {author}
  {\bibfnamefont {S.}~\bibnamefont {Messelt}}, \bibinfo {author} {\bibfnamefont
  {G.~E.}\ \bibnamefont {Mitchell}}, \bibinfo {author} {\bibfnamefont
  {J.}~\bibnamefont {Rekstad}}, \bibinfo {author} {\bibfnamefont
  {A.}~\bibnamefont {Schiller}}, \bibinfo {author} {\bibfnamefont
  {S.}~\bibnamefont {Siem}}, \bibinfo {author} {\bibfnamefont {A.~C.}\
  \bibnamefont {Sunde}}, \bibinfo {author} {\bibfnamefont {A.}~\bibnamefont
  {Voinov}}, \ and\ \bibinfo {author} {\bibfnamefont {S.}~\bibnamefont
  {\O{}deg\aa{}rd}},\ }\href {\doibase 10.1103/PhysRevC.71.044307} {\bibfield
  {journal} {\bibinfo  {journal} {Phys. Rev. C}\ }\textbf {\bibinfo {volume}
  {71}},\ \bibinfo {pages} {044307} (\bibinfo {year} {2005})}\BibitemShut
  {NoStop}%
\bibitem [{\citenamefont {Wiedeking}\ \emph {et~al.}(2012)\citenamefont
  {Wiedeking}, \citenamefont {Bernstein}, \citenamefont {Krti\v{c}ka},
  \citenamefont {Bleuel}, \citenamefont {Allmond}, \citenamefont {Basunia},
  \citenamefont {Harke}, \citenamefont {Fallon}, \citenamefont {Firestone},
  \citenamefont {Goldblum}, \citenamefont {Hatarik}, \citenamefont {Lake},
  \citenamefont {Lee}, \citenamefont {Lesher}, \citenamefont {Paschalis},
  \citenamefont {Petri}, \citenamefont {Phair}, ,\ and\ \citenamefont
  {Scielzo}}]{wiedeking_2012}%
  \BibitemOpen
  \bibfield  {author} {\bibinfo {author} {\bibfnamefont {M.}~\bibnamefont
  {Wiedeking}}, \bibinfo {author} {\bibfnamefont {L.~A.}\ \bibnamefont
  {Bernstein}}, \bibinfo {author} {\bibfnamefont {M.}~\bibnamefont
  {Krti\v{c}ka}}, \bibinfo {author} {\bibfnamefont {D.~L.}\ \bibnamefont
  {Bleuel}}, \bibinfo {author} {\bibfnamefont {J.~M.}\ \bibnamefont {Allmond}},
  \bibinfo {author} {\bibfnamefont {M.~S.}\ \bibnamefont {Basunia}}, \bibinfo
  {author} {\bibfnamefont {J.~T.}\ \bibnamefont {Harke}}, \bibinfo {author}
  {\bibfnamefont {P.}~\bibnamefont {Fallon}}, \bibinfo {author} {\bibfnamefont
  {R.~B.}\ \bibnamefont {Firestone}}, \bibinfo {author} {\bibfnamefont {B.~L.}\
  \bibnamefont {Goldblum}}, \bibinfo {author} {\bibfnamefont {R.}~\bibnamefont
  {Hatarik}}, \bibinfo {author} {\bibfnamefont {P.~T.}\ \bibnamefont {Lake}},
  \bibinfo {author} {\bibfnamefont {I.-Y.}\ \bibnamefont {Lee}}, \bibinfo
  {author} {\bibfnamefont {S.~R.}\ \bibnamefont {Lesher}}, \bibinfo {author}
  {\bibfnamefont {S.}~\bibnamefont {Paschalis}}, \bibinfo {author}
  {\bibfnamefont {M.}~\bibnamefont {Petri}}, \bibinfo {author} {\bibfnamefont
  {L.}~\bibnamefont {Phair}}, , \ and\ \bibinfo {author} {\bibfnamefont
  {N.~D.}\ \bibnamefont {Scielzo}},\ }\href@noop {} {\bibfield  {journal}
  {\bibinfo  {journal} {Phys. Rev. Lett.}\ }\textbf {\bibinfo {volume} {108}},\
  \bibinfo {pages} {162503} (\bibinfo {year} {2012})}\BibitemShut {NoStop}%
\bibitem [{\citenamefont {Larsen}\ \emph {et~al.}(2013)\citenamefont {Larsen},
  \citenamefont {Blasi}, \citenamefont {Bracco}, \citenamefont {Camera},
  \citenamefont {Eriksen}, \citenamefont {G\"orgen}, \citenamefont
  {Guttormsen}, \citenamefont {Hagen}, \citenamefont {Leoni}, \citenamefont
  {Million}, \citenamefont {Nyhus}, \citenamefont {Renstr{\o}m}, \citenamefont
  {Rose}, \citenamefont {Ruud}, \citenamefont {Siem}, \citenamefont {Tornyi},
  \citenamefont {Tveten}, \citenamefont {Voinov}, ,\ and\ \citenamefont
  {Wiedeking}}]{larsen_2013}%
  \BibitemOpen
  \bibfield  {author} {\bibinfo {author} {\bibfnamefont {A.~C.}\ \bibnamefont
  {Larsen}}, \bibinfo {author} {\bibfnamefont {N.}~\bibnamefont {Blasi}},
  \bibinfo {author} {\bibfnamefont {A.}~\bibnamefont {Bracco}}, \bibinfo
  {author} {\bibfnamefont {F.}~\bibnamefont {Camera}}, \bibinfo {author}
  {\bibfnamefont {T.~K.}\ \bibnamefont {Eriksen}}, \bibinfo {author}
  {\bibfnamefont {A.}~\bibnamefont {G\"orgen}}, \bibinfo {author}
  {\bibfnamefont {M.}~\bibnamefont {Guttormsen}}, \bibinfo {author}
  {\bibfnamefont {T.~W.}\ \bibnamefont {Hagen}}, \bibinfo {author}
  {\bibfnamefont {S.}~\bibnamefont {Leoni}}, \bibinfo {author} {\bibfnamefont
  {B.}~\bibnamefont {Million}}, \bibinfo {author} {\bibfnamefont {H.~T.}\
  \bibnamefont {Nyhus}}, \bibinfo {author} {\bibfnamefont {T.}~\bibnamefont
  {Renstr{\o}m}}, \bibinfo {author} {\bibfnamefont {S.~J.}\ \bibnamefont
  {Rose}}, \bibinfo {author} {\bibfnamefont {I.~E.}\ \bibnamefont {Ruud}},
  \bibinfo {author} {\bibfnamefont {S.}~\bibnamefont {Siem}}, \bibinfo {author}
  {\bibfnamefont {T.}~\bibnamefont {Tornyi}}, \bibinfo {author} {\bibfnamefont
  {G.~M.}\ \bibnamefont {Tveten}}, \bibinfo {author} {\bibfnamefont {A.~V.}\
  \bibnamefont {Voinov}}, , \ and\ \bibinfo {author} {\bibfnamefont
  {M.}~\bibnamefont {Wiedeking}},\ }\href@noop {} {\bibfield  {journal}
  {\bibinfo  {journal} {Phys. Rev. Lett.}\ }\textbf {\bibinfo {volume} {111}},\
  \bibinfo {pages} {242504} (\bibinfo {year} {2013})}\BibitemShut {NoStop}%
\bibitem [{\citenamefont {Larsen}\ \emph {et~al.}(2018)\citenamefont {Larsen},
  \citenamefont {Midtb\o{}}, \citenamefont {Guttormsen}, \citenamefont
  {Renstr\o{}m}, \citenamefont {Liddick}, \citenamefont {Spyrou}, \citenamefont
  {Karampagia}, \citenamefont {Brown}, \citenamefont {Achakovskiy},
  \citenamefont {Kamerdzhiev}, \citenamefont {Bleuel}, \citenamefont {Couture},
  \citenamefont {Campo}, \citenamefont {Crider}, \citenamefont {Dombos},
  \citenamefont {Lewis}, \citenamefont {Mosby}, \citenamefont {Naqvi},
  \citenamefont {Perdikakis}, \citenamefont {Prokop}, \citenamefont {Quinn},\
  and\ \citenamefont {Siem}}]{larsen_2018}%
  \BibitemOpen
  \bibfield  {author} {\bibinfo {author} {\bibfnamefont {A.~C.}\ \bibnamefont
  {Larsen}}, \bibinfo {author} {\bibfnamefont {J.~E.}\ \bibnamefont
  {Midtb\o{}}}, \bibinfo {author} {\bibfnamefont {M.}~\bibnamefont
  {Guttormsen}}, \bibinfo {author} {\bibfnamefont {T.}~\bibnamefont
  {Renstr\o{}m}}, \bibinfo {author} {\bibfnamefont {S.~N.}\ \bibnamefont
  {Liddick}}, \bibinfo {author} {\bibfnamefont {A.}~\bibnamefont {Spyrou}},
  \bibinfo {author} {\bibfnamefont {S.}~\bibnamefont {Karampagia}}, \bibinfo
  {author} {\bibfnamefont {B.~A.}\ \bibnamefont {Brown}}, \bibinfo {author}
  {\bibfnamefont {O.}~\bibnamefont {Achakovskiy}}, \bibinfo {author}
  {\bibfnamefont {S.}~\bibnamefont {Kamerdzhiev}}, \bibinfo {author}
  {\bibfnamefont {D.~L.}\ \bibnamefont {Bleuel}}, \bibinfo {author}
  {\bibfnamefont {A.}~\bibnamefont {Couture}}, \bibinfo {author} {\bibfnamefont
  {L.~C.}\ \bibnamefont {Campo}}, \bibinfo {author} {\bibfnamefont {B.~P.}\
  \bibnamefont {Crider}}, \bibinfo {author} {\bibfnamefont {A.~C.}\
  \bibnamefont {Dombos}}, \bibinfo {author} {\bibfnamefont {R.}~\bibnamefont
  {Lewis}}, \bibinfo {author} {\bibfnamefont {S.}~\bibnamefont {Mosby}},
  \bibinfo {author} {\bibfnamefont {F.}~\bibnamefont {Naqvi}}, \bibinfo
  {author} {\bibfnamefont {G.}~\bibnamefont {Perdikakis}}, \bibinfo {author}
  {\bibfnamefont {C.~J.}\ \bibnamefont {Prokop}}, \bibinfo {author}
  {\bibfnamefont {S.~J.}\ \bibnamefont {Quinn}}, \ and\ \bibinfo {author}
  {\bibfnamefont {S.}~\bibnamefont {Siem}},\ }\href {\doibase
  10.1103/PhysRevC.97.054329} {\bibfield  {journal} {\bibinfo  {journal} {Phys.
  Rev. C}\ }\textbf {\bibinfo {volume} {97}},\ \bibinfo {pages} {054329}
  (\bibinfo {year} {2018})}\BibitemShut {NoStop}%
\bibitem [{\citenamefont {Kheswa}\ \emph {et~al.}(2015)\citenamefont {Kheswa},
  \citenamefont {Wiedeking}, \citenamefont {Giacoppo}, \citenamefont {Goriely},
  \citenamefont {Guttormsen}, \citenamefont {Larsen}, \citenamefont {Garrote},
  \citenamefont {Eriksen}, \citenamefont {G{\"o}rgen}, \citenamefont {Hagen},
  \citenamefont {Koehler}, \citenamefont {Klintefjord}, \citenamefont {Nyhus},
  \citenamefont {Papka}, \citenamefont {Renstr{\o}m}, \citenamefont {Rose},
  \citenamefont {Sahin}, \citenamefont {Siem},\ and\ \citenamefont
  {Tornyi}}]{kheswa_2015}%
  \BibitemOpen
  \bibfield  {author} {\bibinfo {author} {\bibfnamefont {B.~V.}\ \bibnamefont
  {Kheswa}}, \bibinfo {author} {\bibfnamefont {M.}~\bibnamefont {Wiedeking}},
  \bibinfo {author} {\bibfnamefont {F.}~\bibnamefont {Giacoppo}}, \bibinfo
  {author} {\bibfnamefont {S.}~\bibnamefont {Goriely}}, \bibinfo {author}
  {\bibfnamefont {M.}~\bibnamefont {Guttormsen}}, \bibinfo {author}
  {\bibfnamefont {A.~C.}\ \bibnamefont {Larsen}}, \bibinfo {author}
  {\bibfnamefont {F.~L.~B.}\ \bibnamefont {Garrote}}, \bibinfo {author}
  {\bibfnamefont {T.~K.}\ \bibnamefont {Eriksen}}, \bibinfo {author}
  {\bibfnamefont {A.}~\bibnamefont {G{\"o}rgen}}, \bibinfo {author}
  {\bibfnamefont {T.~W.}\ \bibnamefont {Hagen}}, \bibinfo {author}
  {\bibfnamefont {P.~E.}\ \bibnamefont {Koehler}}, \bibinfo {author}
  {\bibfnamefont {M.}~\bibnamefont {Klintefjord}}, \bibinfo {author}
  {\bibfnamefont {H.~T.}\ \bibnamefont {Nyhus}}, \bibinfo {author}
  {\bibfnamefont {P.}~\bibnamefont {Papka}}, \bibinfo {author} {\bibfnamefont
  {T.}~\bibnamefont {Renstr{\o}m}}, \bibinfo {author} {\bibfnamefont
  {S.}~\bibnamefont {Rose}}, \bibinfo {author} {\bibfnamefont {E.}~\bibnamefont
  {Sahin}}, \bibinfo {author} {\bibfnamefont {S.}~\bibnamefont {Siem}}, \ and\
  \bibinfo {author} {\bibfnamefont {T.}~\bibnamefont {Tornyi}},\ }\href@noop {}
  {\bibfield  {journal} {\bibinfo  {journal} {Phys. Lett. B}\ }\textbf
  {\bibinfo {volume} {744}},\ \bibinfo {pages} {268} (\bibinfo {year}
  {2015})}\BibitemShut {NoStop}%
\bibitem [{\citenamefont {Naqvi}\ \emph {et~al.}(2019)\citenamefont {Naqvi},
  \citenamefont {Simon}, \citenamefont {Guttormsen}, \citenamefont
  {Schwengner}, \citenamefont {Frauendorf}, \citenamefont {Reingold},
  \citenamefont {Harke}, \citenamefont {Cooper}, \citenamefont {Hughes},
  \citenamefont {Ota},\ and\ \citenamefont {Saastamoinen}}]{navqi_2019}%
  \BibitemOpen
  \bibfield  {author} {\bibinfo {author} {\bibfnamefont {F.}~\bibnamefont
  {Naqvi}}, \bibinfo {author} {\bibfnamefont {A.}~\bibnamefont {Simon}},
  \bibinfo {author} {\bibfnamefont {M.}~\bibnamefont {Guttormsen}}, \bibinfo
  {author} {\bibfnamefont {R.}~\bibnamefont {Schwengner}}, \bibinfo {author}
  {\bibfnamefont {S.}~\bibnamefont {Frauendorf}}, \bibinfo {author}
  {\bibfnamefont {C.~S.}\ \bibnamefont {Reingold}}, \bibinfo {author}
  {\bibfnamefont {J.~T.}\ \bibnamefont {Harke}}, \bibinfo {author}
  {\bibfnamefont {N.}~\bibnamefont {Cooper}}, \bibinfo {author} {\bibfnamefont
  {R.~O.}\ \bibnamefont {Hughes}}, \bibinfo {author} {\bibfnamefont
  {S.}~\bibnamefont {Ota}}, \ and\ \bibinfo {author} {\bibfnamefont
  {A.}~\bibnamefont {Saastamoinen}},\ }\href@noop {} {\bibfield  {journal}
  {\bibinfo  {journal} {Phys. Rev. C}\ }\textbf {\bibinfo {volume} {99}},\
  \bibinfo {pages} {054331} (\bibinfo {year} {2019})}\BibitemShut {NoStop}%
\bibitem [{\citenamefont {Simon}\ \emph {et~al.}(2016)\citenamefont {Simon},
  \citenamefont {Guttormsen}, \citenamefont {Larsen}, \citenamefont {Beausang},
  \citenamefont {Humby}, \citenamefont {Harke}, \citenamefont {Casperson},
  \citenamefont {Hughes}, \citenamefont {Ross}, \citenamefont {Allmond},
  \citenamefont {Chyzh}, \citenamefont {Dag}, \citenamefont {Koglin},
  \citenamefont {McCleskey}, \citenamefont {McCleskey}, \citenamefont {Ota},\
  and\ \citenamefont {Saastamoinen}}]{simon_2016}%
  \BibitemOpen
  \bibfield  {author} {\bibinfo {author} {\bibfnamefont {A.}~\bibnamefont
  {Simon}}, \bibinfo {author} {\bibfnamefont {M.}~\bibnamefont {Guttormsen}},
  \bibinfo {author} {\bibfnamefont {A.~C.}\ \bibnamefont {Larsen}}, \bibinfo
  {author} {\bibfnamefont {C.~W.}\ \bibnamefont {Beausang}}, \bibinfo {author}
  {\bibfnamefont {P.}~\bibnamefont {Humby}}, \bibinfo {author} {\bibfnamefont
  {J.~T.}\ \bibnamefont {Harke}}, \bibinfo {author} {\bibfnamefont {R.~J.}\
  \bibnamefont {Casperson}}, \bibinfo {author} {\bibfnamefont {R.~O.}\
  \bibnamefont {Hughes}}, \bibinfo {author} {\bibfnamefont {T.~J.}\
  \bibnamefont {Ross}}, \bibinfo {author} {\bibfnamefont {J.~M.}\ \bibnamefont
  {Allmond}}, \bibinfo {author} {\bibfnamefont {R.}~\bibnamefont {Chyzh}},
  \bibinfo {author} {\bibfnamefont {M.}~\bibnamefont {Dag}}, \bibinfo {author}
  {\bibfnamefont {J.}~\bibnamefont {Koglin}}, \bibinfo {author} {\bibfnamefont
  {E.}~\bibnamefont {McCleskey}}, \bibinfo {author} {\bibfnamefont
  {M.}~\bibnamefont {McCleskey}}, \bibinfo {author} {\bibfnamefont
  {S.}~\bibnamefont {Ota}}, \ and\ \bibinfo {author} {\bibfnamefont
  {A.}~\bibnamefont {Saastamoinen}},\ }\href@noop {} {\bibfield  {journal}
  {\bibinfo  {journal} {Phys. Rev. C}\ }\textbf {\bibinfo {volume} {93}},\
  \bibinfo {pages} {034303} (\bibinfo {year} {2016})}\BibitemShut {NoStop}%
\bibitem [{\citenamefont {Guttormsen}\ \emph {et~al.}(2022)\citenamefont
  {Guttormsen}, \citenamefont {Ay}, \citenamefont {Ozgur}, \citenamefont
  {Algin}, \citenamefont {Larsen}, \citenamefont {Bello~Garrote}, \citenamefont
  {Berg}, \citenamefont {Crespo~Campo}, \citenamefont {Dahl-Jacobsen},
  \citenamefont {Furmyr}, \citenamefont {Gjestvang}, \citenamefont {G\"orgen},
  \citenamefont {Hagen}, \citenamefont {Ingeberg}, \citenamefont {Kheswa},
  \citenamefont {Kullmann}, \citenamefont {Klintefjord}, \citenamefont
  {Markova}, \citenamefont {Midtb\o{}}, \citenamefont {Modamio}, \citenamefont
  {Paulsen}, \citenamefont {Pedersen}, \citenamefont {Renstr\o{}m},
  \citenamefont {Sahin}, \citenamefont {Siem}, \citenamefont {Tveten},\ and\
  \citenamefont {Wiedeking}}]{guttormsen_2022}%
  \BibitemOpen
  \bibfield  {author} {\bibinfo {author} {\bibfnamefont {M.}~\bibnamefont
  {Guttormsen}}, \bibinfo {author} {\bibfnamefont {K.~O.}\ \bibnamefont {Ay}},
  \bibinfo {author} {\bibfnamefont {M.}~\bibnamefont {Ozgur}}, \bibinfo
  {author} {\bibfnamefont {E.}~\bibnamefont {Algin}}, \bibinfo {author}
  {\bibfnamefont {A.~C.}\ \bibnamefont {Larsen}}, \bibinfo {author}
  {\bibfnamefont {F.~L.}\ \bibnamefont {Bello~Garrote}}, \bibinfo {author}
  {\bibfnamefont {H.~C.}\ \bibnamefont {Berg}}, \bibinfo {author}
  {\bibfnamefont {L.}~\bibnamefont {Crespo~Campo}}, \bibinfo {author}
  {\bibfnamefont {T.}~\bibnamefont {Dahl-Jacobsen}}, \bibinfo {author}
  {\bibfnamefont {F.~W.}\ \bibnamefont {Furmyr}}, \bibinfo {author}
  {\bibfnamefont {D.}~\bibnamefont {Gjestvang}}, \bibinfo {author}
  {\bibfnamefont {A.}~\bibnamefont {G\"orgen}}, \bibinfo {author}
  {\bibfnamefont {T.~W.}\ \bibnamefont {Hagen}}, \bibinfo {author}
  {\bibfnamefont {V.~W.}\ \bibnamefont {Ingeberg}}, \bibinfo {author}
  {\bibfnamefont {B.~V.}\ \bibnamefont {Kheswa}}, \bibinfo {author}
  {\bibfnamefont {I.~K.~B.}\ \bibnamefont {Kullmann}}, \bibinfo {author}
  {\bibfnamefont {M.}~\bibnamefont {Klintefjord}}, \bibinfo {author}
  {\bibfnamefont {M.}~\bibnamefont {Markova}}, \bibinfo {author} {\bibfnamefont
  {J.~E.}\ \bibnamefont {Midtb\o{}}}, \bibinfo {author} {\bibfnamefont
  {V.}~\bibnamefont {Modamio}}, \bibinfo {author} {\bibfnamefont
  {W.}~\bibnamefont {Paulsen}}, \bibinfo {author} {\bibfnamefont {L.~G.}\
  \bibnamefont {Pedersen}}, \bibinfo {author} {\bibfnamefont {T.}~\bibnamefont
  {Renstr\o{}m}}, \bibinfo {author} {\bibfnamefont {E.}~\bibnamefont {Sahin}},
  \bibinfo {author} {\bibfnamefont {S.}~\bibnamefont {Siem}}, \bibinfo {author}
  {\bibfnamefont {G.~M.}\ \bibnamefont {Tveten}}, \ and\ \bibinfo {author}
  {\bibfnamefont {M.}~\bibnamefont {Wiedeking}},\ }\href {\doibase
  10.1103/PhysRevC.106.034314} {\bibfield  {journal} {\bibinfo  {journal}
  {Phys. Rev. C}\ }\textbf {\bibinfo {volume} {106}},\ \bibinfo {pages}
  {034314} (\bibinfo {year} {2022})}\BibitemShut {NoStop}%
\bibitem [{\citenamefont {Jones}\ \emph {et~al.}(2018)\citenamefont {Jones},
  \citenamefont {Macchiavelli}, \citenamefont {Wiedeking}, \citenamefont
  {Bernstein}, \citenamefont {Crawford}, \citenamefont {Campbell},
  \citenamefont {Clark}, \citenamefont {Cromaz}, \citenamefont {Fallon},
  \citenamefont {Lee}, \citenamefont {Salathe}, \citenamefont {Wiens},
  \citenamefont {Ayangeakaa}, \citenamefont {Bleuel}, \citenamefont {Bottoni},
  \citenamefont {Carpenter}, \citenamefont {Davids}, \citenamefont {Elson},
  \citenamefont {G\"orgen}, \citenamefont {Guttormsen}, \citenamefont
  {Janssens}, \citenamefont {Kinnison}, \citenamefont {Kirsch}, \citenamefont
  {Larsen}, \citenamefont {Lauritsen}, \citenamefont {Reviol}, \citenamefont
  {Sarantites}, \citenamefont {Siem}, \citenamefont {Voinov},\ and\
  \citenamefont {Zhu}}]{jones_2018}%
  \BibitemOpen
  \bibfield  {author} {\bibinfo {author} {\bibfnamefont {M.~D.}\ \bibnamefont
  {Jones}}, \bibinfo {author} {\bibfnamefont {A.~O.}\ \bibnamefont
  {Macchiavelli}}, \bibinfo {author} {\bibfnamefont {M.}~\bibnamefont
  {Wiedeking}}, \bibinfo {author} {\bibfnamefont {L.~A.}\ \bibnamefont
  {Bernstein}}, \bibinfo {author} {\bibfnamefont {H.~L.}\ \bibnamefont
  {Crawford}}, \bibinfo {author} {\bibfnamefont {C.~M.}\ \bibnamefont
  {Campbell}}, \bibinfo {author} {\bibfnamefont {R.~M.}\ \bibnamefont {Clark}},
  \bibinfo {author} {\bibfnamefont {M.}~\bibnamefont {Cromaz}}, \bibinfo
  {author} {\bibfnamefont {P.}~\bibnamefont {Fallon}}, \bibinfo {author}
  {\bibfnamefont {I.~Y.}\ \bibnamefont {Lee}}, \bibinfo {author} {\bibfnamefont
  {M.}~\bibnamefont {Salathe}}, \bibinfo {author} {\bibfnamefont
  {A.}~\bibnamefont {Wiens}}, \bibinfo {author} {\bibfnamefont {A.~D.}\
  \bibnamefont {Ayangeakaa}}, \bibinfo {author} {\bibfnamefont {D.~L.}\
  \bibnamefont {Bleuel}}, \bibinfo {author} {\bibfnamefont {S.}~\bibnamefont
  {Bottoni}}, \bibinfo {author} {\bibfnamefont {M.~P.}\ \bibnamefont
  {Carpenter}}, \bibinfo {author} {\bibfnamefont {H.~M.}\ \bibnamefont
  {Davids}}, \bibinfo {author} {\bibfnamefont {J.}~\bibnamefont {Elson}},
  \bibinfo {author} {\bibfnamefont {A.}~\bibnamefont {G\"orgen}}, \bibinfo
  {author} {\bibfnamefont {M.}~\bibnamefont {Guttormsen}}, \bibinfo {author}
  {\bibfnamefont {R.~V.~F.}\ \bibnamefont {Janssens}}, \bibinfo {author}
  {\bibfnamefont {J.~E.}\ \bibnamefont {Kinnison}}, \bibinfo {author}
  {\bibfnamefont {L.}~\bibnamefont {Kirsch}}, \bibinfo {author} {\bibfnamefont
  {A.~C.}\ \bibnamefont {Larsen}}, \bibinfo {author} {\bibfnamefont
  {T.}~\bibnamefont {Lauritsen}}, \bibinfo {author} {\bibfnamefont
  {W.}~\bibnamefont {Reviol}}, \bibinfo {author} {\bibfnamefont {D.~G.}\
  \bibnamefont {Sarantites}}, \bibinfo {author} {\bibfnamefont
  {S.}~\bibnamefont {Siem}}, \bibinfo {author} {\bibfnamefont {A.~V.}\
  \bibnamefont {Voinov}}, \ and\ \bibinfo {author} {\bibfnamefont
  {S.}~\bibnamefont {Zhu}},\ }\href {\doibase 10.1103/PhysRevC.97.024327}
  {\bibfield  {journal} {\bibinfo  {journal} {Phys. Rev. C}\ }\textbf {\bibinfo
  {volume} {97}},\ \bibinfo {pages} {024327} (\bibinfo {year}
  {2018})}\BibitemShut {NoStop}%
\bibitem [{\citenamefont {Larsen}\ and\ \citenamefont
  {Goriely}(2010)}]{larsen_2010}%
  \BibitemOpen
  \bibfield  {author} {\bibinfo {author} {\bibfnamefont {A.~C.}\ \bibnamefont
  {Larsen}}\ and\ \bibinfo {author} {\bibfnamefont {S.}~\bibnamefont
  {Goriely}},\ }\href {\doibase 10.1103/PhysRevC.82.014318} {\bibfield
  {journal} {\bibinfo  {journal} {Phys. Rev. C}\ }\textbf {\bibinfo {volume}
  {82}},\ \bibinfo {pages} {014318} (\bibinfo {year} {2010})}\BibitemShut
  {NoStop}%
\bibitem [{\citenamefont {Loens}\ \emph {et~al.}(2012)\citenamefont {Loens},
  \citenamefont {Langanke}, \citenamefont {Mart\'inez-Pinedo},\ and\
  \citenamefont {Sieja}}]{loens_2012}%
  \BibitemOpen
  \bibfield  {author} {\bibinfo {author} {\bibfnamefont {H.}~\bibnamefont
  {Loens}}, \bibinfo {author} {\bibfnamefont {K.}~\bibnamefont {Langanke}},
  \bibinfo {author} {\bibfnamefont {G.}~\bibnamefont {Mart\'inez-Pinedo}}, \
  and\ \bibinfo {author} {\bibfnamefont {K.}~\bibnamefont {Sieja}},\
  }\href@noop {} {\bibfield  {journal} {\bibinfo  {journal} {Eur. Phys. J. A}\
  }\textbf {\bibinfo {volume} {48}},\ \bibinfo {pages} {34} (\bibinfo {year}
  {2012})}\BibitemShut {NoStop}%
\bibitem [{\citenamefont {Schwengner}\ \emph {et~al.}(2013)\citenamefont
  {Schwengner}, \citenamefont {Frauendorf},\ and\ \citenamefont
  {Larsen}}]{schwengner_2013}%
  \BibitemOpen
  \bibfield  {author} {\bibinfo {author} {\bibfnamefont {R.}~\bibnamefont
  {Schwengner}}, \bibinfo {author} {\bibfnamefont {S.}~\bibnamefont
  {Frauendorf}}, \ and\ \bibinfo {author} {\bibfnamefont {A.~C.}\ \bibnamefont
  {Larsen}},\ }\href {\doibase 10.1103/PhysRevLett.111.232504} {\bibfield
  {journal} {\bibinfo  {journal} {Phys. Rev. Lett.}\ }\textbf {\bibinfo
  {volume} {111}},\ \bibinfo {pages} {232504} (\bibinfo {year}
  {2013})}\BibitemShut {NoStop}%
\bibitem [{\citenamefont {Brown}\ and\ \citenamefont
  {Larsen}(2014)}]{brown_2014}%
  \BibitemOpen
  \bibfield  {author} {\bibinfo {author} {\bibfnamefont {B.~A.}\ \bibnamefont
  {Brown}}\ and\ \bibinfo {author} {\bibfnamefont {A.~C.}\ \bibnamefont
  {Larsen}},\ }\href {\doibase 10.1103/PhysRevLett.113.252502} {\bibfield
  {journal} {\bibinfo  {journal} {Phys. Rev. Lett.}\ }\textbf {\bibinfo
  {volume} {113}},\ \bibinfo {pages} {252502} (\bibinfo {year}
  {2014})}\BibitemShut {NoStop}%
\bibitem [{\citenamefont {Schwengner}\ \emph {et~al.}(2017)\citenamefont
  {Schwengner}, \citenamefont {Frauendorf},\ and\ \citenamefont
  {Brown}}]{schwengner_2017}%
  \BibitemOpen
  \bibfield  {author} {\bibinfo {author} {\bibfnamefont {R.}~\bibnamefont
  {Schwengner}}, \bibinfo {author} {\bibfnamefont {S.}~\bibnamefont
  {Frauendorf}}, \ and\ \bibinfo {author} {\bibfnamefont {B.~A.}\ \bibnamefont
  {Brown}},\ }\href {\doibase 10.1103/PhysRevLett.118.092502} {\bibfield
  {journal} {\bibinfo  {journal} {Phys. Rev. Lett.}\ }\textbf {\bibinfo
  {volume} {118}},\ \bibinfo {pages} {092502} (\bibinfo {year}
  {2017})}\BibitemShut {NoStop}%
\bibitem [{\citenamefont {Sieja}(2017)}]{sieja_2017}%
  \BibitemOpen
  \bibfield  {author} {\bibinfo {author} {\bibfnamefont {K.}~\bibnamefont
  {Sieja}},\ }\href@noop {} {\bibfield  {journal} {\bibinfo  {journal} {Phys.
  Rev. Lett.}\ }\textbf {\bibinfo {volume} {119}},\ \bibinfo {pages} {052502}
  (\bibinfo {year} {2017})}\BibitemShut {NoStop}%
\bibitem [{\citenamefont {Karampagia}\ \emph {et~al.}(2017)\citenamefont
  {Karampagia}, \citenamefont {Brown},\ and\ \citenamefont
  {Zelevinsky}}]{karampagia_2017}%
  \BibitemOpen
  \bibfield  {author} {\bibinfo {author} {\bibfnamefont {S.}~\bibnamefont
  {Karampagia}}, \bibinfo {author} {\bibfnamefont {B.~A.}\ \bibnamefont
  {Brown}}, \ and\ \bibinfo {author} {\bibfnamefont {V.}~\bibnamefont
  {Zelevinsky}},\ }\href@noop {} {\bibfield  {journal} {\bibinfo  {journal}
  {Phys. Rev. C}\ }\textbf {\bibinfo {volume} {95}},\ \bibinfo {pages} {024322}
  (\bibinfo {year} {2017})}\BibitemShut {NoStop}%
\bibitem [{\citenamefont {Sieja}(2018)}]{sieja_2018}%
  \BibitemOpen
  \bibfield  {author} {\bibinfo {author} {\bibfnamefont {K.}~\bibnamefont
  {Sieja}},\ }\href {\doibase 10.1103/PhysRevC.98.064312} {\bibfield  {journal}
  {\bibinfo  {journal} {Phys. Rev. C}\ }\textbf {\bibinfo {volume} {98}},\
  \bibinfo {pages} {064312} (\bibinfo {year} {2018})}\BibitemShut {NoStop}%
\bibitem [{\citenamefont {Litvinova}\ and\ \citenamefont
  {Belov}(2013)}]{litvinova_2013}%
  \BibitemOpen
  \bibfield  {author} {\bibinfo {author} {\bibfnamefont {E.}~\bibnamefont
  {Litvinova}}\ and\ \bibinfo {author} {\bibfnamefont {N.}~\bibnamefont
  {Belov}},\ }\href@noop {} {\bibfield  {journal} {\bibinfo  {journal} {Phys.
  Rev. C}\ }\textbf {\bibinfo {volume} {88}},\ \bibinfo {pages} {031302(R)}
  (\bibinfo {year} {2013})}\BibitemShut {NoStop}%
\bibitem [{\citenamefont {Mitdb{\o}}\ \emph {et~al.}(2018)\citenamefont
  {Mitdb{\o}}, \citenamefont {Larsen}, \citenamefont {Renstr{\o}m},
  \citenamefont {Garrote},\ and\ \citenamefont {Lime}}]{mitdbo_2018}%
  \BibitemOpen
  \bibfield  {author} {\bibinfo {author} {\bibfnamefont {J.~E.}\ \bibnamefont
  {Mitdb{\o}}}, \bibinfo {author} {\bibfnamefont {A.~C.}\ \bibnamefont
  {Larsen}}, \bibinfo {author} {\bibfnamefont {T.}~\bibnamefont {Renstr{\o}m}},
  \bibinfo {author} {\bibfnamefont {F.~L.~B.}\ \bibnamefont {Garrote}}, \ and\
  \bibinfo {author} {\bibfnamefont {E.}~\bibnamefont {Lime}},\ }\href@noop {}
  {\bibfield  {journal} {\bibinfo  {journal} {Phys. Rev. C}\ }\textbf {\bibinfo
  {volume} {98}},\ \bibinfo {pages} {064321} (\bibinfo {year}
  {2018})}\BibitemShut {NoStop}%
\bibitem [{\citenamefont {Johnson}\ \emph {et~al.}(1992)\citenamefont
  {Johnson}, \citenamefont {Koonin}, \citenamefont {Lang},\ and\ \citenamefont
  {Ormand}}]{johnson_1992}%
  \BibitemOpen
  \bibfield  {author} {\bibinfo {author} {\bibfnamefont {C.~W.}\ \bibnamefont
  {Johnson}}, \bibinfo {author} {\bibfnamefont {S.~E.}\ \bibnamefont {Koonin}},
  \bibinfo {author} {\bibfnamefont {G.~H.}\ \bibnamefont {Lang}}, \ and\
  \bibinfo {author} {\bibfnamefont {W.~E.}\ \bibnamefont {Ormand}},\
  }\href@noop {} {\bibfield  {journal} {\bibinfo  {journal} {Phys. Rev. Lett.}\
  }\textbf {\bibinfo {volume} {69}},\ \bibinfo {pages} {3157} (\bibinfo {year}
  {1992})}\BibitemShut {NoStop}%
\bibitem [{\citenamefont {Alhassid}\ \emph {et~al.}(1994)\citenamefont
  {Alhassid}, \citenamefont {Dean}, \citenamefont {Koonin}, \citenamefont
  {Lang},\ and\ \citenamefont {Ormand}}]{alhassid_1994}%
  \BibitemOpen
  \bibfield  {author} {\bibinfo {author} {\bibfnamefont {Y.}~\bibnamefont
  {Alhassid}}, \bibinfo {author} {\bibfnamefont {D.~J.}\ \bibnamefont {Dean}},
  \bibinfo {author} {\bibfnamefont {S.~E.}\ \bibnamefont {Koonin}}, \bibinfo
  {author} {\bibfnamefont {G.}~\bibnamefont {Lang}}, \ and\ \bibinfo {author}
  {\bibfnamefont {W.~E.}\ \bibnamefont {Ormand}},\ }\href {\doibase
  10.1103/PhysRevLett.72.613} {\bibfield  {journal} {\bibinfo  {journal} {Phys.
  Rev. Lett.}\ }\textbf {\bibinfo {volume} {72}},\ \bibinfo {pages} {613}
  (\bibinfo {year} {1994})}\BibitemShut {NoStop}%
\bibitem [{\citenamefont {Nakada}\ and\ \citenamefont
  {Alhassid}(1997)}]{nakada_1997}%
  \BibitemOpen
  \bibfield  {author} {\bibinfo {author} {\bibfnamefont {H.}~\bibnamefont
  {Nakada}}\ and\ \bibinfo {author} {\bibfnamefont {Y.}~\bibnamefont
  {Alhassid}},\ }\href@noop {} {\bibfield  {journal} {\bibinfo  {journal}
  {Phys. Rev. Lett.}\ }\textbf {\bibinfo {volume} {79}},\ \bibinfo {pages}
  {2939} (\bibinfo {year} {1997})}\BibitemShut {NoStop}%
\bibitem [{\citenamefont {Nakada}\ and\ \citenamefont
  {Alhassid}(1998)}]{nakada_1998}%
  \BibitemOpen
  \bibfield  {author} {\bibinfo {author} {\bibfnamefont {H.}~\bibnamefont
  {Nakada}}\ and\ \bibinfo {author} {\bibfnamefont {Y.}~\bibnamefont
  {Alhassid}},\ }\href {\doibase https://doi.org/10.1016/S0370-2693(98)00911-3}
  {\bibfield  {journal} {\bibinfo  {journal} {Physics Letters B}\ }\textbf
  {\bibinfo {volume} {436}},\ \bibinfo {pages} {231} (\bibinfo {year}
  {1998})}\BibitemShut {NoStop}%
\bibitem [{\citenamefont {Ormand}(1997)}]{ormand_1997}%
  \BibitemOpen
  \bibfield  {author} {\bibinfo {author} {\bibfnamefont {W.~E.}\ \bibnamefont
  {Ormand}},\ }\href@noop {} {\bibfield  {journal} {\bibinfo  {journal} {Phys.
  Rev. C}\ }\textbf {\bibinfo {volume} {56}},\ \bibinfo {pages} {R1678}
  (\bibinfo {year} {1997})}\BibitemShut {NoStop}%
\bibitem [{\citenamefont {Langanke}(1998)}]{langanke_1998}%
  \BibitemOpen
  \bibfield  {author} {\bibinfo {author} {\bibfnamefont {K.}~\bibnamefont
  {Langanke}},\ }\href@noop {} {\bibfield  {journal} {\bibinfo  {journal}
  {Phys. Lett. B}\ }\textbf {\bibinfo {volume} {438}},\ \bibinfo {pages} {235}
  (\bibinfo {year} {1998})}\BibitemShut {NoStop}%
\bibitem [{\citenamefont {Alhassid}\ \emph {et~al.}(1999)\citenamefont
  {Alhassid}, \citenamefont {Liu},\ and\ \citenamefont
  {Nakada}}]{alhassid_1999}%
  \BibitemOpen
  \bibfield  {author} {\bibinfo {author} {\bibfnamefont {Y.}~\bibnamefont
  {Alhassid}}, \bibinfo {author} {\bibfnamefont {S.}~\bibnamefont {Liu}}, \
  and\ \bibinfo {author} {\bibfnamefont {H.}~\bibnamefont {Nakada}},\
  }\href@noop {} {\bibfield  {journal} {\bibinfo  {journal} {Phys. Rev. Lett.}\
  }\textbf {\bibinfo {volume} {83}},\ \bibinfo {pages} {4265} (\bibinfo {year}
  {1999})}\BibitemShut {NoStop}%
\bibitem [{\citenamefont {Alhassid}\ \emph {et~al.}(2007)\citenamefont
  {Alhassid}, \citenamefont {Liu},\ and\ \citenamefont
  {Nakada}}]{alhassid_2007}%
  \BibitemOpen
  \bibfield  {author} {\bibinfo {author} {\bibfnamefont {Y.}~\bibnamefont
  {Alhassid}}, \bibinfo {author} {\bibfnamefont {S.}~\bibnamefont {Liu}}, \
  and\ \bibinfo {author} {\bibfnamefont {H.}~\bibnamefont {Nakada}},\
  }\href@noop {} {\bibfield  {journal} {\bibinfo  {journal} {Phys. Rev. Lett.}\
  }\textbf {\bibinfo {volume} {99}},\ \bibinfo {pages} {162504} (\bibinfo
  {year} {2007})}\BibitemShut {NoStop}%
\bibitem [{\citenamefont {Alhassid}\ \emph {et~al.}(2008)\citenamefont
  {Alhassid}, \citenamefont {Fang},\ and\ \citenamefont
  {Nakada}}]{alhassid_2008}%
  \BibitemOpen
  \bibfield  {author} {\bibinfo {author} {\bibfnamefont {Y.}~\bibnamefont
  {Alhassid}}, \bibinfo {author} {\bibfnamefont {L.}~\bibnamefont {Fang}}, \
  and\ \bibinfo {author} {\bibfnamefont {H.}~\bibnamefont {Nakada}},\
  }\href@noop {} {\bibfield  {journal} {\bibinfo  {journal} {Phys. Rev. Lett.}\
  }\textbf {\bibinfo {volume} {101}},\ \bibinfo {pages} {082501} (\bibinfo
  {year} {2008})}\BibitemShut {NoStop}%
\bibitem [{\citenamefont {Alhassid}()}]{alhassid_2017_book}%
  \BibitemOpen
  \bibfield  {author} {\bibinfo {author} {\bibfnamefont {Y.}~\bibnamefont
  {Alhassid}},\ }\href@noop {} {\emph {\bibinfo {title} {Emergent Phenomena in
  Atomic Nuclei from Large-Scale Modeling: a Symmetry-Guided Perspective}}},\
  edited by\ \bibinfo {editor} {\bibfnamefont {K.~D.}\ \bibnamefont {Launey}}\
  (\bibinfo  {publisher} {World Scientific, Singapore, 2017})\ pp.\ \bibinfo
  {pages} {267--298}\BibitemShut {NoStop}%
\bibitem [{\citenamefont {Gubernatis}\ \emph {et~al.}(1991)\citenamefont
  {Gubernatis}, \citenamefont {Jarrell}, \citenamefont {Silver},\ and\
  \citenamefont {Sivia}}]{gubernatis_1991}%
  \BibitemOpen
  \bibfield  {author} {\bibinfo {author} {\bibfnamefont {J.~E.}\ \bibnamefont
  {Gubernatis}}, \bibinfo {author} {\bibfnamefont {M.}~\bibnamefont {Jarrell}},
  \bibinfo {author} {\bibfnamefont {R.~N.}\ \bibnamefont {Silver}}, \ and\
  \bibinfo {author} {\bibfnamefont {D.~S.}\ \bibnamefont {Sivia}},\ }\href
  {\doibase 10.1103/PhysRevB.44.6011} {\bibfield  {journal} {\bibinfo
  {journal} {Phys. Rev. B}\ }\textbf {\bibinfo {volume} {44}},\ \bibinfo
  {pages} {6011} (\bibinfo {year} {1991})}\BibitemShut {NoStop}%
\bibitem [{\citenamefont {Jarrell}\ and\ \citenamefont
  {Gubernatis}(1996)}]{jarrell_1996}%
  \BibitemOpen
  \bibfield  {author} {\bibinfo {author} {\bibfnamefont {M.}~\bibnamefont
  {Jarrell}}\ and\ \bibinfo {author} {\bibfnamefont {J.~E.}\ \bibnamefont
  {Gubernatis}},\ }\href@noop {} {\bibfield  {journal} {\bibinfo  {journal}
  {Phys. Rep.}\ }\textbf {\bibinfo {volume} {269}},\ \bibinfo {pages} {133}
  (\bibinfo {year} {1996})}\BibitemShut {NoStop}%
\bibitem [{\citenamefont {Gubernatis}\ and\ \citenamefont
  {Werner}(2016)}]{gubernatis_book}%
  \BibitemOpen
  \bibfield  {author} {\bibinfo {author} {\bibfnamefont {J.~E.}\ \bibnamefont
  {Gubernatis}}\ and\ \bibinfo {author} {\bibfnamefont {P.}~\bibnamefont
  {Werner}},\ }\href@noop {} {\emph {\bibinfo {title} {Quantum Monte Carlo
  methods: algorithms for lattice models}}}\ (\bibinfo  {publisher} {Cambridge
  University Press},\ \bibinfo {address} {Cambridge},\ \bibinfo {year}
  {2016})\BibitemShut {NoStop}%
\bibitem [{\citenamefont {Fanto}\ and\ \citenamefont
  {Alhassid}(2024)}]{fanto_2024}%
  \BibitemOpen
  \bibfield  {author} {\bibinfo {author} {\bibfnamefont {P.}~\bibnamefont
  {Fanto}}\ and\ \bibinfo {author} {\bibfnamefont {Y.}~\bibnamefont
  {Alhassid}},\ }\href {\doibase 10.1103/PhysRevC.109.L031302} {\bibfield
  {journal} {\bibinfo  {journal} {Phys. Rev. C}\ }\textbf {\bibinfo {volume}
  {109}},\ \bibinfo {pages} {L031302} (\bibinfo {year} {2024})}\BibitemShut
  {NoStop}%
\bibitem [{\citenamefont {\"Ozen}\ \emph {et~al.}(2013)\citenamefont {\"Ozen},
  \citenamefont {Alhassid},\ and\ \citenamefont {Nakada}}]{ozen_2013}%
  \BibitemOpen
  \bibfield  {author} {\bibinfo {author} {\bibfnamefont {C.}~\bibnamefont
  {\"Ozen}}, \bibinfo {author} {\bibfnamefont {Y.}~\bibnamefont {Alhassid}}, \
  and\ \bibinfo {author} {\bibfnamefont {H.}~\bibnamefont {Nakada}},\ }\href
  {\doibase 10.1103/PhysRevLett.110.042502} {\bibfield  {journal} {\bibinfo
  {journal} {Phys. Rev. Lett.}\ }\textbf {\bibinfo {volume} {110}},\ \bibinfo
  {pages} {042502} (\bibinfo {year} {2013})}\BibitemShut {NoStop}%
\bibitem [{\citenamefont {M\"uhlschlegel}\ \emph {et~al.}(1972)\citenamefont
  {M\"uhlschlegel}, \citenamefont {Scalapino},\ and\ \citenamefont
  {Denton}}]{muhlschlegel_1972}%
  \BibitemOpen
  \bibfield  {author} {\bibinfo {author} {\bibfnamefont {B.}~\bibnamefont
  {M\"uhlschlegel}}, \bibinfo {author} {\bibfnamefont {D.~J.}\ \bibnamefont
  {Scalapino}}, \ and\ \bibinfo {author} {\bibfnamefont {R.}~\bibnamefont
  {Denton}},\ }\href {\doibase 10.1103/PhysRevB.6.1767} {\bibfield  {journal}
  {\bibinfo  {journal} {Phys. Rev. B}\ }\textbf {\bibinfo {volume} {6}},\
  \bibinfo {pages} {1767} (\bibinfo {year} {1972})}\BibitemShut {NoStop}%
\bibitem [{\citenamefont {Alhassid}\ and\ \citenamefont
  {Zingman}(1984)}]{zingman_1984}%
  \BibitemOpen
  \bibfield  {author} {\bibinfo {author} {\bibfnamefont {Y.}~\bibnamefont
  {Alhassid}}\ and\ \bibinfo {author} {\bibfnamefont {J.}~\bibnamefont
  {Zingman}},\ }\href {\doibase 10.1103/PhysRevC.30.684} {\bibfield  {journal}
  {\bibinfo  {journal} {Phys. Rev. C}\ }\textbf {\bibinfo {volume} {30}},\
  \bibinfo {pages} {684} (\bibinfo {year} {1984})}\BibitemShut {NoStop}%
\bibitem [{\citenamefont {Lauritzen}\ \emph {et~al.}(1988)\citenamefont
  {Lauritzen}, \citenamefont {Arve},\ and\ \citenamefont
  {Bertsch}}]{lauritzen_1988}%
  \BibitemOpen
  \bibfield  {author} {\bibinfo {author} {\bibfnamefont {B.}~\bibnamefont
  {Lauritzen}}, \bibinfo {author} {\bibfnamefont {P.}~\bibnamefont {Arve}}, \
  and\ \bibinfo {author} {\bibfnamefont {G.~F.}\ \bibnamefont {Bertsch}},\
  }\href@noop {} {\bibfield  {journal} {\bibinfo  {journal} {Phys. Rev. Lett.}\
  }\textbf {\bibinfo {volume} {61}},\ \bibinfo {pages} {2835} (\bibinfo {year}
  {1988})}\BibitemShut {NoStop}%
\bibitem [{\citenamefont {Arve}\ \emph {et~al.}(1988)\citenamefont {Arve},
  \citenamefont {Bertsch}, \citenamefont {Lauritzen},\ and\ \citenamefont
  {Puddu}}]{arve_1988}%
  \BibitemOpen
  \bibfield  {author} {\bibinfo {author} {\bibfnamefont {P.}~\bibnamefont
  {Arve}}, \bibinfo {author} {\bibfnamefont {G.}~\bibnamefont {Bertsch}},
  \bibinfo {author} {\bibfnamefont {B.}~\bibnamefont {Lauritzen}}, \ and\
  \bibinfo {author} {\bibfnamefont {G.}~\bibnamefont {Puddu}},\ }\href
  {\doibase https://doi.org/10.1016/0003-4916(88)90235-7} {\bibfield  {journal}
  {\bibinfo  {journal} {Annals of Physics}\ }\textbf {\bibinfo {volume}
  {183}},\ \bibinfo {pages} {309} (\bibinfo {year} {1988})}\BibitemShut
  {NoStop}%
\bibitem [{\citenamefont {Attias}\ and\ \citenamefont
  {Alhassid}(1997)}]{attias_1997}%
  \BibitemOpen
  \bibfield  {author} {\bibinfo {author} {\bibfnamefont {H.}~\bibnamefont
  {Attias}}\ and\ \bibinfo {author} {\bibfnamefont {Y.}~\bibnamefont
  {Alhassid}},\ }\href@noop {} {\bibfield  {journal} {\bibinfo  {journal}
  {Nucl. Phys. A}\ }\textbf {\bibinfo {volume} {652}},\ \bibinfo {pages} {565}
  (\bibinfo {year} {1997})}\BibitemShut {NoStop}%
\bibitem [{\citenamefont {Frosini}\ \emph {et~al.}(2024)\citenamefont
  {Frosini}, \citenamefont {Ryssens},\ and\ \citenamefont
  {Sieja}}]{frosini_2023}%
  \BibitemOpen
  \bibfield  {author} {\bibinfo {author} {\bibfnamefont {M.}~\bibnamefont
  {Frosini}}, \bibinfo {author} {\bibfnamefont {W.}~\bibnamefont {Ryssens}}, \
  and\ \bibinfo {author} {\bibfnamefont {K.}~\bibnamefont {Sieja}},\ }\href
  {\doibase 10.1103/PhysRevC.110.014307} {\bibfield  {journal} {\bibinfo
  {journal} {Phys. Rev. C}\ }\textbf {\bibinfo {volume} {110}},\ \bibinfo
  {pages} {014307} (\bibinfo {year} {2024})}\BibitemShut {NoStop}%
\bibitem [{\citenamefont {Hubbard}(1959)}]{hubbard}%
  \BibitemOpen
  \bibfield  {author} {\bibinfo {author} {\bibfnamefont {J.}~\bibnamefont
  {Hubbard}},\ }\href@noop {} {\bibfield  {journal} {\bibinfo  {journal} {Phys.
  Rev. Lett.}\ }\textbf {\bibinfo {volume} {3}},\ \bibinfo {pages} {77}
  (\bibinfo {year} {1959})}\BibitemShut {NoStop}%
\bibitem [{\citenamefont {Stratonovich}(1957)}]{stratonovich_1957}%
  \BibitemOpen
  \bibfield  {author} {\bibinfo {author} {\bibfnamefont {R.~L.}\ \bibnamefont
  {Stratonovich}},\ }\href@noop {} {\bibfield  {journal} {\bibinfo  {journal}
  {Dokl. Akad. Nauk SSSR}\ }\textbf {\bibinfo {volume} {115}},\ \bibinfo
  {pages} {1097} (\bibinfo {year} {1957})}\BibitemShut {NoStop}%
\bibitem [{\citenamefont {Stratonovich}(1958)}]{stratonovich_1958}%
  \BibitemOpen
  \bibfield  {author} {\bibinfo {author} {\bibfnamefont {R.~L.}\ \bibnamefont
  {Stratonovich}},\ }\href@noop {} {\bibfield  {journal} {\bibinfo  {journal}
  {Sov. Phys. Dokl.}\ }\textbf {\bibinfo {volume} {2}},\ \bibinfo {pages} {416}
  (\bibinfo {year} {1958})}\BibitemShut {NoStop}%
\bibitem [{\citenamefont {Koonin}\ \emph {et~al.}(1997)\citenamefont {Koonin},
  \citenamefont {Dean},\ and\ \citenamefont {Langanke}}]{koonin_1997}%
  \BibitemOpen
  \bibfield  {author} {\bibinfo {author} {\bibfnamefont {S.~E.}\ \bibnamefont
  {Koonin}}, \bibinfo {author} {\bibfnamefont {D.~J.}\ \bibnamefont {Dean}}, \
  and\ \bibinfo {author} {\bibfnamefont {K.}~\bibnamefont {Langanke}},\
  }\href@noop {} {\bibfield  {journal} {\bibinfo  {journal} {Phys. Rep.}\
  }\textbf {\bibinfo {volume} {278}},\ \bibinfo {pages} {1} (\bibinfo {year}
  {1997})}\BibitemShut {NoStop}%
\bibitem [{\citenamefont {Alhassid}\ and\ \citenamefont
  {Bush}(1990)}]{alhassid_1990}%
  \BibitemOpen
  \bibfield  {author} {\bibinfo {author} {\bibfnamefont {Y.}~\bibnamefont
  {Alhassid}}\ and\ \bibinfo {author} {\bibfnamefont {B.}~\bibnamefont
  {Bush}},\ }\href {\doibase https://doi.org/10.1016/0375-9474(90)90087-3}
  {\bibfield  {journal} {\bibinfo  {journal} {Nuclear Physics A}\ }\textbf
  {\bibinfo {volume} {509}},\ \bibinfo {pages} {461} (\bibinfo {year}
  {1990})}\BibitemShut {NoStop}%
\bibitem [{\citenamefont {Bryan}(1990)}]{bryan_1990}%
  \BibitemOpen
  \bibfield  {author} {\bibinfo {author} {\bibfnamefont {R.~K.}\ \bibnamefont
  {Bryan}},\ }\href@noop {} {\bibfield  {journal} {\bibinfo  {journal} {Eur.
  Biophys. J.}\ }\textbf {\bibinfo {volume} {18}},\ \bibinfo {pages} {165}
  (\bibinfo {year} {1990})}\BibitemShut {NoStop}%
\bibitem [{\citenamefont {Ryssens}\ and\ \citenamefont
  {Alhassid}()}]{ryssens_2024}%
  \BibitemOpen
  \bibfield  {author} {\bibinfo {author} {\bibfnamefont {W.}~\bibnamefont
  {Ryssens}}\ and\ \bibinfo {author} {\bibfnamefont {Y.}~\bibnamefont
  {Alhassid}},\ }\href@noop {} {\bibinfo  {journal} {Solution of the
  finite-temperature (q)rpa equations for shell-model Hamiltonians: HF-SHELL
  v2, in preparation}\ }\BibitemShut {NoStop}%
\bibitem [{\citenamefont {Ryssens}\ and\ \citenamefont
  {Alhassid}(2021)}]{ryssens_2021}%
  \BibitemOpen
\bibfield  {journal} {  }\bibfield  {author} {\bibinfo {author} {\bibfnamefont
  {W.}~\bibnamefont {Ryssens}}\ and\ \bibinfo {author} {\bibfnamefont
  {Y.}~\bibnamefont {Alhassid}},\ }\href@noop {} {\bibfield  {journal}
  {\bibinfo  {journal} {Eur. Phys. J. A}\ }\textbf {\bibinfo {volume} {57}},\
  \bibinfo {pages} {76} (\bibinfo {year} {2021})}\BibitemShut {NoStop}%
\bibitem [{\citenamefont {Fanto}\ and\ \citenamefont
  {Alhassid}(2021)}]{fanto_2021}%
  \BibitemOpen
  \bibfield  {author} {\bibinfo {author} {\bibfnamefont {P.}~\bibnamefont
  {Fanto}}\ and\ \bibinfo {author} {\bibfnamefont {Y.}~\bibnamefont
  {Alhassid}},\ }\href@noop {} {\bibfield  {journal} {\bibinfo  {journal}
  {Phys. Rev. C}\ }\textbf {\bibinfo {volume} {103}},\ \bibinfo {pages}
  {064310} (\bibinfo {year} {2021})}\BibitemShut {NoStop}%
\bibitem [{\citenamefont {Rossignoli}\ and\ \citenamefont
  {Ring}(1998)}]{rossignoli_1998}%
  \BibitemOpen
  \bibfield  {author} {\bibinfo {author} {\bibfnamefont {R.}~\bibnamefont
  {Rossignoli}}\ and\ \bibinfo {author} {\bibfnamefont {P.}~\bibnamefont
  {Ring}},\ }\href {\doibase https://doi.org/10.1016/S0375-9474(98)00815-X}
  {\bibfield  {journal} {\bibinfo  {journal} {Nuclear Physics A}\ }\textbf
  {\bibinfo {volume} {633}},\ \bibinfo {pages} {613} (\bibinfo {year}
  {1998})}\BibitemShut {NoStop}%
\bibitem [{\citenamefont {Bohr}\ and\ \citenamefont
  {Mottelson}()}]{bohr_1969_book}%
  \BibitemOpen
  \bibfield  {author} {\bibinfo {author} {\bibfnamefont {A.}~\bibnamefont
  {Bohr}}\ and\ \bibinfo {author} {\bibfnamefont {B.~R.}\ \bibnamefont
  {Mottelson}},\ }\href@noop {} {\emph {\bibinfo {title} {Nuclear
  Structure}}},\ Vol.~\bibinfo {volume} {1}\ (\bibinfo  {publisher} {Benjamin,
  New York, 1969})\BibitemShut {NoStop}%
\bibitem [{\citenamefont {Alhassid}\ \emph {et~al.}(1996)\citenamefont
  {Alhassid}, \citenamefont {Bertsch}, \citenamefont {Dean},\ and\
  \citenamefont {Koonin}}]{alhassid_1996}%
  \BibitemOpen
  \bibfield  {author} {\bibinfo {author} {\bibfnamefont {Y.}~\bibnamefont
  {Alhassid}}, \bibinfo {author} {\bibfnamefont {G.~F.}\ \bibnamefont
  {Bertsch}}, \bibinfo {author} {\bibfnamefont {D.~J.}\ \bibnamefont {Dean}}, \
  and\ \bibinfo {author} {\bibfnamefont {S.~E.}\ \bibnamefont {Koonin}},\
  }\href@noop {} {\bibfield  {journal} {\bibinfo  {journal} {Phys. Rev. Lett.}\
  }\textbf {\bibinfo {volume} {77}},\ \bibinfo {pages} {1444} (\bibinfo {year}
  {1996})}\BibitemShut {NoStop}%
\bibitem [{\citenamefont {Krti\v{c}ka}\ and\ \citenamefont
  {Be\v{c}v\`a\v{r}}(2011)}]{krticka_2011}%
  \BibitemOpen
  \bibfield  {author} {\bibinfo {author} {\bibfnamefont {M.}~\bibnamefont
  {Krti\v{c}ka}}\ and\ \bibinfo {author} {\bibfnamefont {F.}~\bibnamefont
  {Be\v{c}v\`a\v{r}}},\ }\href@noop {} {\bibfield  {journal} {\bibinfo
  {journal} {Int. J. Mod. Phys. E}\ }\textbf {\bibinfo {volume} {20}},\
  \bibinfo {pages} {488} (\bibinfo {year} {2011})}\BibitemShut {NoStop}%
\bibitem [{dal()}]{dallas}%
  \BibitemOpen
  \href@noop {} {}\bibinfo {note} {D. DeMartini, private
  communication.}\BibitemShut {Stop}%
\bibitem [{\citenamefont {Guttormsen}\ \emph {et~al.}(2021)\citenamefont
  {Guttormsen}, \citenamefont {Alhassid}, \citenamefont {Ryssens},
  \citenamefont {Ay}, \citenamefont {Ozgur}, \citenamefont {Algin},
  \citenamefont {Larsen}, \citenamefont {{Bello Garrote}}, \citenamefont
  {{Crespo Campo}}, \citenamefont {Dahl-Jacobsen}, \citenamefont {Görgen},
  \citenamefont {Hagen}, \citenamefont {Ingeberg}, \citenamefont {Kheswa},
  \citenamefont {Klintefjord}, \citenamefont {Midtbø}, \citenamefont
  {Modamio}, \citenamefont {Renstrøm}, \citenamefont {Sahin}, \citenamefont
  {Siem}, \citenamefont {Tveten},\ and\ \citenamefont
  {Zeiser}}]{guttormsen_2021}%
  \BibitemOpen
  \bibfield  {author} {\bibinfo {author} {\bibfnamefont {M.}~\bibnamefont
  {Guttormsen}}, \bibinfo {author} {\bibfnamefont {Y.}~\bibnamefont
  {Alhassid}}, \bibinfo {author} {\bibfnamefont {W.}~\bibnamefont {Ryssens}},
  \bibinfo {author} {\bibfnamefont {K.}~\bibnamefont {Ay}}, \bibinfo {author}
  {\bibfnamefont {M.}~\bibnamefont {Ozgur}}, \bibinfo {author} {\bibfnamefont
  {E.}~\bibnamefont {Algin}}, \bibinfo {author} {\bibfnamefont
  {A.}~\bibnamefont {Larsen}}, \bibinfo {author} {\bibfnamefont
  {F.}~\bibnamefont {{Bello Garrote}}}, \bibinfo {author} {\bibfnamefont
  {L.}~\bibnamefont {{Crespo Campo}}}, \bibinfo {author} {\bibfnamefont
  {T.}~\bibnamefont {Dahl-Jacobsen}}, \bibinfo {author} {\bibfnamefont
  {A.}~\bibnamefont {Görgen}}, \bibinfo {author} {\bibfnamefont
  {T.}~\bibnamefont {Hagen}}, \bibinfo {author} {\bibfnamefont
  {V.}~\bibnamefont {Ingeberg}}, \bibinfo {author} {\bibfnamefont
  {B.}~\bibnamefont {Kheswa}}, \bibinfo {author} {\bibfnamefont
  {M.}~\bibnamefont {Klintefjord}}, \bibinfo {author} {\bibfnamefont
  {J.}~\bibnamefont {Midtbø}}, \bibinfo {author} {\bibfnamefont
  {V.}~\bibnamefont {Modamio}}, \bibinfo {author} {\bibfnamefont
  {T.}~\bibnamefont {Renstrøm}}, \bibinfo {author} {\bibfnamefont
  {E.}~\bibnamefont {Sahin}}, \bibinfo {author} {\bibfnamefont
  {S.}~\bibnamefont {Siem}}, \bibinfo {author} {\bibfnamefont {G.}~\bibnamefont
  {Tveten}}, \ and\ \bibinfo {author} {\bibfnamefont {F.}~\bibnamefont
  {Zeiser}},\ }\href {\doibase https://doi.org/10.1016/j.physletb.2021.136206}
  {\bibfield  {journal} {\bibinfo  {journal} {Physics Letters B}\ }\textbf
  {\bibinfo {volume} {816}},\ \bibinfo {pages} {136206} (\bibinfo {year}
  {2021})}\BibitemShut {NoStop}%
\end{thebibliography}

\end{document}